 \documentclass[twocolumn]{aastex631}

\shorttitle{Hydrogen burning of millinovae} 
\shortauthors{Hachisu \& Kato}

\begin{document}

%% LaTeX will automatically break titles if they run longer than
%% one line. However, you may use \\ to force a line break if
%% you desire.

% should be written in CAPITAL
\title{A demarcation criterion for hydrogen burning of millinovae}

%% Use \author, \affil, and the \and command to format
%% author and affiliation information.
%% Note that \email has replaced the old \authoremail command
%% from AASTeX v4.0. You can use \email to mark an email address
%% anywhere in the paper, not just in the front matter.
%% As in the title, use \\ to force line breaks.

\author[0000-0002-0884-7404]{Izumi Hachisu}
\affil{Department of Earth Science and Astronomy, College of Arts and
Sciences,  University of Tokyo, 3-8-1 Komaba, Meguro-ku,
Tokyo 153-8902, Japan}
\email{izumi.hachisu@outlook.jp}

%\and

\author[0000-0002-8522-8033]{Mariko Kato} 
\affil{Department of Astronomy, Keio University, Hiyoshi, Yokohama
  223-8521, Japan}
% \email{mariko.kato@hc.st.keio.ac.jp}

%\author{Hideyuki Saio}
%\affil{Astronomical Institute, Graduate School of Science,
%    Tohoku University, Sendai 980-8578, Japan}
% \email{saio@astr.tohoku.ac.jp}

%% Notice that each of these authors has alternate affiliations, which
%% are identified by the \altaffilmark after each name.  Specify alternate
%% affiliation information with \altaffiltext, with one command per each
%% affiliation.

%% Mark off your abstract in the ``abstract'' environment. In the manuscript
%% style, abstract will output a Received/Accepted line after the
%% title and affiliation information. No date will appear since the author
%% does not have this information. The dates will be filled in by the
%% editorial office after submission.

\begin{abstract} 
Millinovae are a new class of transient supersoft X-ray sources with no clear
signature of mass ejection. They show similar triangle shapes of $V/I$ band
light curves with thousand times fainter peaks than typical classical novae.
Maccarone et al. regarded the prototype millinova, ASASSN-16oh, 
as a dwarf nova and interpreted the supersoft X-rays to originate from 
an accretion belt on a white dwarf (WD). 
Kato et al. proposed a nova model induced by a high-rate 
mass-accretion during a dwarf nova outburst; 
the X-rays originate from the photosphere
of a hydrogen-burning hot WD whereas the $V/I$ band photons are from the
irradiated accretion disk. 
Because each peak brightness differs largely from millinova to millinova,
we suspect that not all the millinova candidates host a hydrogen burning WD.
Based on the light curve analysis of the classical nova KT Eri that has
a bright disk,  
we find that the disk is more than two magnitudes brighter when the disk
is irradiated by the hydrogen burning WD than when not irradiated.
We present the demarcation criterion
for hydrogen burning to be $I_{\rm q} - I_{\rm max} > 2.2$, 
where $I_q$ and $I_{\rm max}$ are the $I$ magnitudes in quiescence
and at maximum light, respectively.  Among many candidates, this requirement
is satisfied with the two millinovae in which soft X-rays were detected. 
%%% up to 6 keywords
\end{abstract}

\section{Introduction}
\label{sec_introduction}

A millinova is a transient supersoft X-ray source ($L_X\gtrsim 10^{35}$ erg
s$^{-1}$) with no clear
signature of mass ejection.  The outburst shape of its optical light curve
is almost a symmetric triangle, the peak of which
are thousand times fainter than those of typical classical novae.
Such a new class of cataclysmic variables were dubbed millinovae
by \citet{mro24ks}.

\begin{figure*}
%\epsscale{0.8}
%%\plotone{as16oh_vi_linear.epsi}
\plotone{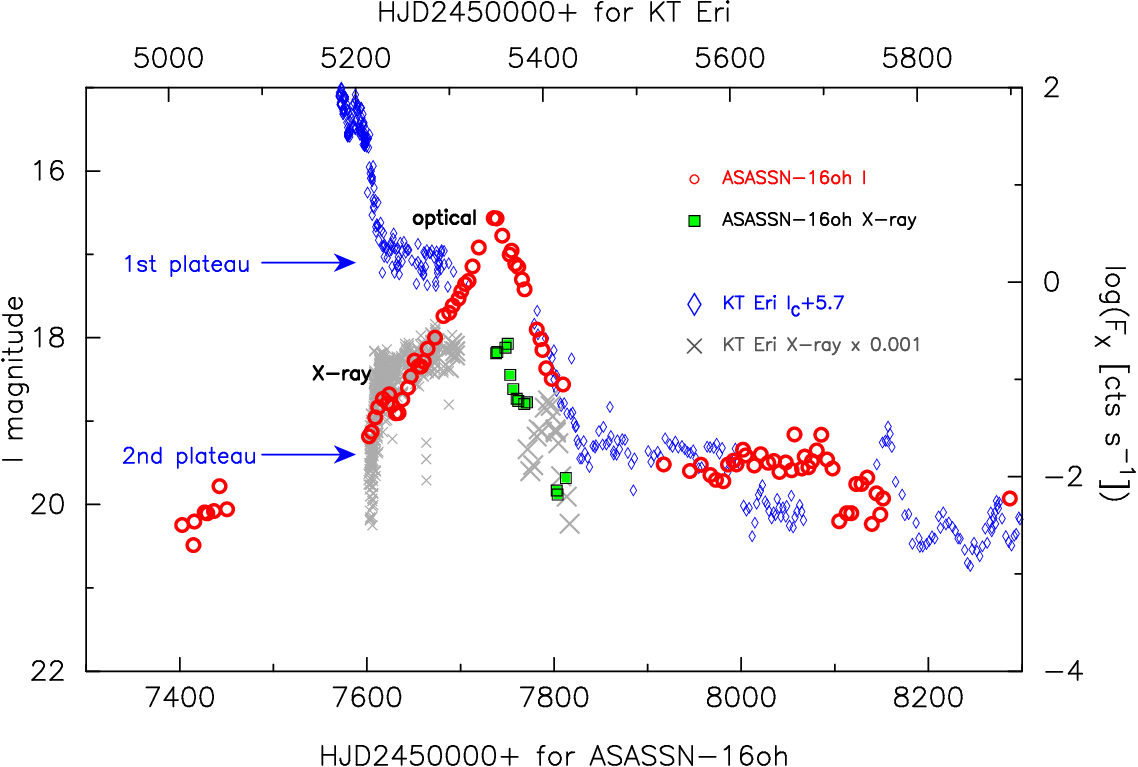}
\caption{
The $I$ light curve of ASASSN-16oh as well as the soft X-ray 
count rates observed with the Swift/XRT.  The data are taken from
the OGLE IV \citep{uda15ss} and the Swift website \citep{eva09}.
The $I_{\rm C}$ and Swift/XRT data of the Galactic classical nova KT Eri
are overplotted at the distance of the Small Magellanic Cloud (SMC),
the data of which are the same as those in \citet{hac25kw}.  The brightnesses
of the first and second plateau phases of KT Eri are indicated
by the blue arrows. 
\label{as16oh_vi_linear}}
\end{figure*}

A representative example of millinovae is ASASSN-16oh, 
which is a peculiar transient supersoft X-ray source 
in the Small Magellanic Cloud (SMC)
with no mass ejection signature \citep{mac19}.  The $I$
light curve of the outburst shows a triangle shape, as
plotted in Figure \ref{as16oh_vi_linear}. 
This figure also shows the  X-ray count rates observed with 
the Swift/XRT (0.3--10.0 keV). 
The peak count rate is converted to the X-ray luminosity of
$L_X\sim 1\times 10^{37}$ erg s$^{-1}$ \citep{mac19}. 

%\placefigure
%Fig.1 

A nova is a thermonuclear runaway 
event on a mass-accreting white dwarf (WD).  
We show the optical and X-ray light curves of the Galactic classical nova
KT Eri in Figures \ref{as16oh_vi_linear} and \ref{kt_eri_bvi_light_curves}.
Novae usually brighten up by $\Delta V\sim 10$ mag or more in a few or tens 
days and decline in a few months or years.
After hydrogen ignites on the WD, the hydrogen-rich envelope expands
to a giant star size and emits strong winds \citep[e.g.][for a recent fully
self-consistent nova model]{kat22sha}.  
The nova brightness is dominated by free-free emission from 
optically thin ejecta outside the photosphere (see Figure
\ref{kt_eri_bvi_light_curves}b).
Thus, the peak optical brightness depends on the 
maximum wind mass-loss rate \citep[e.g.,][]{hac06kb, hac15k}. Due to 
strong mass loss, the envelope loses its mass and the photosphere 
gradually shrinks and the photospheric temperature increases.  
Eventually the nova enters a supersoft X-ray source (SSS) phase
after a large part of the initial envelope mass is lost. 
The envelope mass decreases further due to nuclear burning 
after the winds stop.  
Hydrogen burning ends when the envelope mass reaches a critical value.
As a definition, a nova accompanies strong winds (mass-ejection) 
in the optical bright phase.  The $V/I$ light curve shapes and peak $V/I$
brightnesses of KT Eri are typical in fast novae and very different from
those of millinovae.  There seem no common properties between ASASSN-16oh
and KT Eri except showing a SSS phase and having a bright disk
during the outburst.

A dwarf nova is an rapid accretion event in an accretion disk, the mass
accretion of which is temporarily enhanced by a thermal instability
of an accretion disk \citep[e.g. ][for a review]{osa96}.
If the instability develops from the outer region (outside-in outburst),
the brightness rapidly rises toward maximum.
On the other hand, when the outburst is inside-out (the instability starts
from inside of the disk), the rise is slow and the light curve is more
symmetric (triangle) \citep{las01}. 
Dwarf novae usually brighten up to $M_{V,\rm peak}\sim +6$--$+0$ 
(or by $\Delta V\sim 2$--$9$ mag) 
in a day or so and stays at the brightness from a few days to a few weeks.
The peak brightness depends on the orbital period; the longer
the orbital period, the brighter the $M_{V,\rm peak}$ \citep{pat11}.  
However, no bright SSS phase ($L_X \gtrsim 1\times 10^{35}$ erg s$^{-1}$)
has ever been observed in dwarf nova outbursts
\citep[typically $L_X= 10^{29}$--$10^{32}$
erg s$^{-1}$; ][]{rod24es, schwope24kk}. 

ASASSN-16oh was discovered by the All Sky Automated Survey for Supernovae
(ASASSN) at $V= 16.9$ \citep{jha16}.  Then, the brightness reaches $M_V= -2.3$,
where we assume that the distance modulus is $\mu_0\equiv (m-M)_0=18.9$
and the absorption in the $V$ band is $A_V= 0.13$ toward ASASSN-16oh
\citep{jha16, mro16}.  The Optical Gravitational 
Lensing Experiment IV (OGLE-IV) \citep{uda15ss} data show
that the object is an irregular variable for several years 
with the quiescent luminosity of $I=20.3$ and $V=21.1$ \citep{mac19}.

The $V$ and $I$ light curves of ASASSN-16oh show a resemblance 
with those of long orbital-period dwarf novae such as V1017~Sgr 
in the peak brightness, outburst amplitude, and timescales. 
The orbital period of V1017~Sgr is $P_{\rm orb}= 5.786$~days
\citep{salazar17}.  Therefore, \citet{mac19} expected the orbital 
period of ASASSN-16oh to be several days. \citet{raj17cb} reported
two peaks (0.1925 and 5.6619 days) in the periodgram of 
\ion{He}{2}$\lambda 4686$ spectroscopic velocity data.  
It is in favor of the longer $P_{\rm orb}= 5.6619$ days from
the resemblance with V1017 Sgr.

It has been argued whether or not soft X-rays in millinovae come from
a hydrogen burning WD like in the SSS phase of a nova.
In the present paper, we find that the classical nova KT Eri is one of
the best examples for us to draw such a demarcation line for hydrogen
burning in the color-magnitude diagram.
In Section \ref{models_millinova}, we summarize the previous studies 
on the origin of supersoft X-rays of ASASSN-16oh.
In Section \ref{brightness_irradiated_viscous_disk}, we explain
physical properties of KT Eri and give the outburst evolution track
in the color-magnitude diagram.  We also describe
the brightness difference between the irradiated disk and viscous
heating disk.  Section \ref{demarcation_hydrogen_burning} presents
the demarcation line in the color-magnitude diagram 
and apply it to the 29 millinova candidates proposed by \citet{mro24ks}.
Discussion and conclusions follow in Section \ref{discussion} and
\ref{conclusions}, respectively.

\section{The nature of ASASSN-16\lowercase{oh}}
\label{models_millinova}

\subsection{Maccarone et al.'s model}
\label{maccarone_model}

\citet{mac19} concluded that ASASSN-16oh is an accretion event
like a dwarf nova but unlikely a classical nova
(thermonuclear runaway event), mainly because 
(1) the optical spectra show a very narrow width of
\ion{He}{2} emission lines with no signature of mass-ejection, 
(2) very slow rise ($\sim 200$ days) to the optical maximum compared
with those of classical novae (a few days), 
(3) rather faint peak $V$ magnitude of $M_{V, {\rm max}} \sim -2.3$
compared with those of recurrent novae 
\citep[e.g., the 1 year recurrence period nova
M31N~2008-12a of $M_{V, {\rm max}}= -7.2$,][]{dar15, hen18}.

Only the concern is the origin of supersoft X-rays because  
no bright SSS phase ($L_X \gtrsim 1\times 10^{35}$ erg s$^{-1}$)
has been detected in dwarf nova outbursts.
\citet{mac19} interpreted that the supersoft X-rays could originate from 
a spreading layer around the equatorial accretion belt. 
This model requires a very high mass accretion rate of 
$\dot{M}_{\rm acc}\sim 3\times 10^{-7}~M_\sun$~yr$^{-1}$ 
($2 \times 10^{19}$~g~s$^{-1}$) on a very massive WD of $1.3~M_\sun$ 
to support high temperatures and fluxes for supersoft X-ray. 

The hot spreading layer of the accretion belt has not yet been 
studied in detail. If a disk around a massive WD with high mass-accretion
rates always accompanies a hot spreading layer, 
such supersoft X-rays could be observed in recurrent novae, that is,
in the quiescent phase of U Sco 
\citep[$M_{\rm WD} \sim 1.37~M_\sun$ with $\dot M_{\rm acc} 
\sim 2.5 \times 10^{-7}~M_\sun$yr$^{-1}$,][]{hkkm00} or RS Oph
\citep[$\sim 1.35~M_\sun$ with $\sim 2 \times 10^{-7}~M_\sun$
yr$^{-1}$,][]{hac07kl},  
but not yet detected \citep{kat20sh}.
Because ASASSN-16oh is the first dwarf nova with soft X-ray 
($L_X\sim 1\times 10^{37}$ erg s$^{-1}$) detection,  
its origin of X-rays is particularly interesting. 

\subsection{Hillman et al.'s model}
\label{hillman_model}

\citet{hil19} proposed another interpretation of ASASSN-16oh, 
a thermonuclear runaway event. 
Their models are $M_{\rm WD}=1.1~M_\sun$ WD 
with the mass accretion rate of 
$\dot M_{\rm acc}= (3.5 - 5) \times10^{-7}~M_\sun$ yr$^{-1}$. 
No mass ejection occurs because the shell flash is very weak. 
They calculated $V/I$ light curves assuming 
their $V/I$ photons are emitted from the hot WD surface as well as X-rays. 
Their $V/I$ light curves show rapid-rise and slow-decline shape,  
but the decline slopes are 10 times slower than the observed light curves.
They did not present the X-ray light curves. 
 
\citet{kat20sh} calculated the shell flash models with 
the same parameters as Hillman et al.'s and confirmed that 
a hot WD with a surface temperature $T > 400,000$~K
does not emit many low energy $V/I$ band photons.
\citet{kat20sh} obtained the $V$ band light curves that show 
a good agreement with Hillman et al.'s, and conclude that 
\citet{hil19} adopted a distance modulus $\mu_V \equiv (m-M)_V = 13.0$ 
rather than the canonical value of 19.0 mag to the SMC. 
To summarize, Hillman et al.'s light curve is fainter than ASASSN-16oh 
by 6 mag and decays 10 times slower.

%Fig.2mari  
\begin{figure*}
%\epsscale{0.35}
%\plotone{f2d.eps}
\gridline{\fig{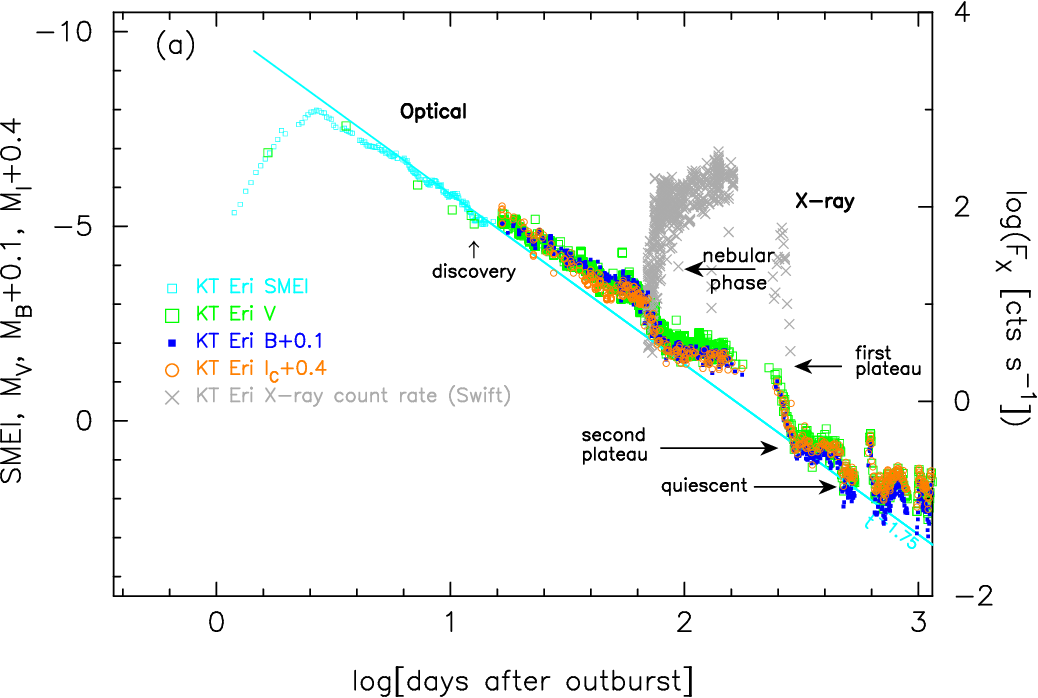}{0.65\textwidth}{}
          }
\gridline{\fig{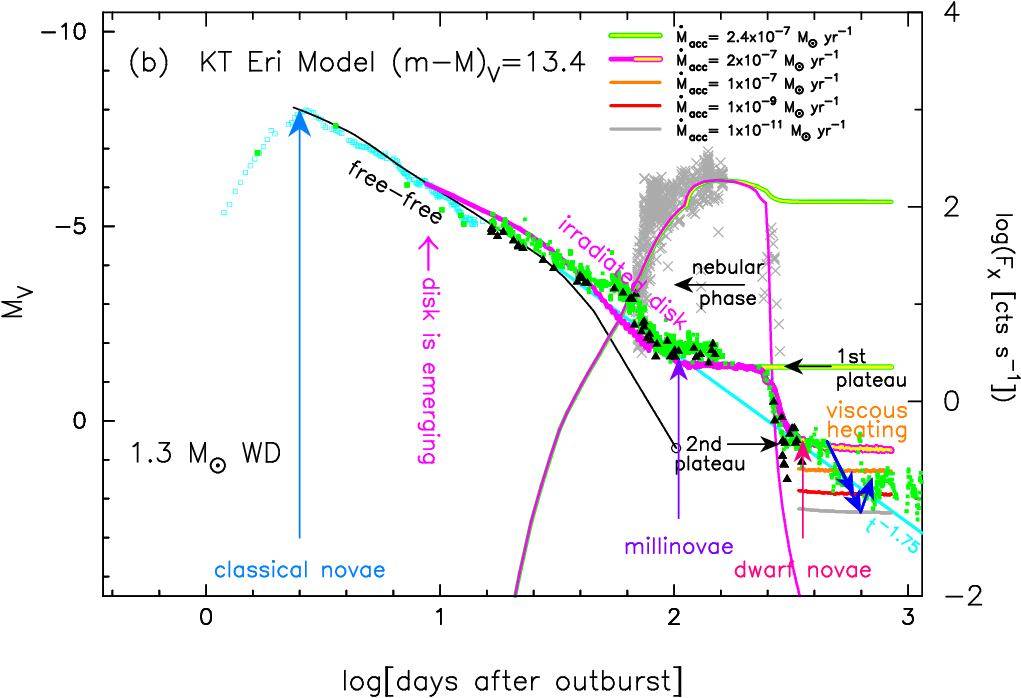}{0.65\textwidth}{}
          }
\caption{(a) 
The $B,V,I_{\rm C},$ and SMEI magnitudes of KT Eri against a logarithmic time.
The $B,I_{\rm C}$ data are taken from AAVSO, VSOLJ, and SMARTS \citep{wal12bt}.
The outburst day of $t_{\rm OB}=$JD 2,455,147.5 is assumed to be
2.7 days before optical (SMEI) maximum.
The SMEI,$V,y$ data are the same as those in \citet{hac25kw}.
The Swift X-ray (0.3--10.0 keV) count rates are also added \citep[taken
from the Swift website; ][]{eva09}.
The global decay trend is described by the universal
decline law of $L_V\propto t^{-1.75}$ \citep[straight cyan line
labeled $t^{-1.75}$:][]{hac06kb, hac23k}.  The X-ray bright phase
(day 68 -- 280) is called the supersoft X-ray source (SSS) phase.
(b) Our model light curves are overplotted on the SMEI, $V$, and $y$
(filled black triangles) light curves.  The black line denotes the
$L_{V,\rm ff,WD}+L_{V,\rm ph,WD}$ model light curve while the thick
magenta line corresponds to the  $L_{V,\rm ff,WD}+L_{V,\rm ph,WD}+
L_{V,\rm ph,disk}+L_{V,\rm ph,comp}$ model light curve for the mass-accretion
rate of $\dot{M}_{\rm acc}= 2\times 10^{-7} ~M_\sun$ yr$^{-1}$.
The other thick colored lines represent different mass-accretion rates.
See the main text for other symbols.
These model light curves are essentially the same as those in \citet{hac25kw}.
\label{kt_eri_bvi_light_curves}}
\end{figure*}

\subsection{Kato et al.'s model}
\label{kato_model}

Assuming that X-rays and $V/I$ photons are emitted from 
different regions as interpreted by \citet{mac19}, 
\citet{kat20sh} proposed a weak shell flash model in which they attributed 
the X-ray emission to the nova SSS phase. 
The main differences from Hillman et al.'s model are:
(1) The $V/I$ photons originate from the irradiated accretion disk and 
companion star because the WD is too hot to emit much $V/I$ photons.  
(2) The WD mass is much more massive ($1.32~M_\sun$) than in
Hillman et al.'s model ($1.1~M_\sun$) to explain the observed fast decay 
in the X-ray light curve.  
(3) The hydrogen shell flash was triggered by a massive mass-inflow 
during a dwarf nova outburst. 

\citet{kat20sh} presented an X-ray light curve 
model of a $1.32~M_\sun$ WD (with a low metallicity of $Z=0.001$) 
that reproduces the temporal decay of the observed X-rays. 
They concluded that ASASSN-16oh is a dwarf-nova that triggers
a nova (hydrogen-burning) with no mass ejection. They further suggested that,
in their model, the WD mass increases steadily 
because of no mass-ejection during the flash, thus,
such a massive mass-increasing WD in a low $Z$ environment
could be a new kind of candidates for Type Ia supernova progenitors.

\subsection{Interpretation on the nature of ASASSN-16oh}

From the above three models, we may conclude that 
(1) the prototype millinova ASASSN-16oh hosts a massive WD, and
(2) its $V/I$ photons mainly come from the (irradiated or viscous heating)
accretion disk.
(3) The X-rays originate from the surface of hydrogen-burning WD or
from the hot accretion belt on the WD.
Because the total of 29 millinovae candidates are reported by 
\citet{mro24ks},
it is critically important to find the demarcation line that easily
tells us whether or not hydrogen burns on the WD.

\section{Brightnesses of irradiated disks and viscous heating disks}
\label{brightness_irradiated_viscous_disk}

When hydrogen burning starts on the WD, the photospheric temperature 
becomes high enough to emit supersoft X-ray photons and the luminosity
increases closely to the Eddington limit.  The photospheric surfaces of
the disk and companion star are irradiated by the hot and bright WD.
The $V/I$ magnitudes of the irradiated disk and companion star  
could become much brighter than those of non-irradiated disk and
companion star even though the disk itself brightens up with viscous
heating. 

We found a good example for comparison 
between irradiated and non-irradiated disk and companion star.
Figure \ref{kt_eri_bvi_light_curves}a shows 
the optical light curves of the classical nova KT Eri.
There are two plateau phases. 
The first plateau corresponds to the SSS phase, 
i.e., the WD emits supersoft X-rays. 
The optical or NIR brightness is governed mainly by the brightness of 
the accretion disk irradiated by the central hot WD. 
The absolute $I$ magnitude of the irradiated disk 
is converted to be about $M_I \sim -2$ at the KT Eri distance modulus of 
$(m-M)_I= 13.3$ \citep{hac25kw}. 
The brightness of the second plateau is explained with the viscous
heating disk because hydrogen burning has already ended 
in this stage \citep[see ][for details]{hac25kw}.

The viscous-heating disk is two or three magnitudes fainter than 
the irradiated disk.  In the second plateau of KT Eri, \citet{hac25kw}
assumed the mass accretion rate to be as high as 
$\dot{M}_{\rm acc}\sim 2\times 10^{-7} ~M_\sun$ yr$^{-1}$,
which is close to the steady hydrogen burning rate of
$\dot{M}_{\rm acc,steady}\approx 2.4\times 10^{-7} ~M_\sun$ yr$^{-1}$
for a $1.3 ~M_\sun$ WD.
If we assume a lower accretion rate, the brightness of the second
plateau becomes fainter as shown in Figure \ref{kt_eri_bvi_light_curves}b. 
If we assume a lager mass accretion rate such as
$\dot{M}_{\rm acc}\gtrsim 2.4\times 10^{-7} ~M_\sun$yr$^{-1}$ 
hydrogen burns steadily and never stops. 
Thus, we may regard that the brightness of the second plateau 
in KT Eri is close to an upper limit for
the brightnesses of viscous heating accretion disks. 

In what follows,
we first describe the $V/I$ light curves of the classical
nova KT Eri and then compare KT Eri with ASASSN-16oh.

%Fig.3
%\placefigure{kt_eri_config}

\begin{figure}
%%%\epsscale{1.0}
\epsscale{1.15}
%%%\epsscale{0.75}
\plotone{f3.eps}
%%%\plottwo{f1a.eps}{f1b.eps}
%%%\plotone{kt_eri_v339_del_v_logscale_no3.epsi}
\caption{
The binary configuration of KT~Eri based on the light curve 
analysis by \citet{hac25kw}.
(a) In the wind phase. The disk surface is blown in the winds and 
its photosphere is slightly expanding.
(b) After the wind stops, the disk photosphere shrinks to a normal size.
The disk and companion star are irradiated by the supersoft X-ray photons 
from the central hot WD.  Such irradiation effects are all included in the
calculation of the $V$ light curve.
The masses of the WD and Roche-lobe-filling companion star
are $1.3 ~M_\sun$ and $1.0 ~M_\sun$, respectively.
The orbital period is $P_{\rm orb}= 2.616$ days.
The inclination angle of the binary is $i=41.2\arcdeg$.
See \citet{hac01kb} for the calculation methods of $V$ light curves.
\label{kt_eri_config}}
\end{figure}

\subsection{Accretion disk in KT Eri}

KT Eri is a fast nova that went into outburst in 2009. 
Unlike typical classical novae, the orbital period is as long as 2.616 days
\citep{schaefer22wh}. Therefore, it could have a large size accretion disk.

Figure \ref{kt_eri_bvi_light_curves}a shows the $B$, $V$, and $I_{\rm C}$
light curves of KT Eri.  These three light curves almost overlap if we
shift down the $B$ and $I_{\rm C}$ light curves by 0.1 and 0.4 mag,
respectively.  This means that the continuum flux dominates the spectra
with no significant contribution of emission lines
to the $BVI_{\rm C}$ bands even in the nebular phase.
This is not a typical property of classical novae. 
The optical light curves decay along with the universal decline law
of $L_V \propto t^{-1.75}$ (thick cyan line labeled $t^{-1.75}$
in Figure \ref{kt_eri_bvi_light_curves}a),
which is a typical property of classical novae \citep{hac06kb, hac23k}.
Here, $t$ is the time from the outburst and $L_V$ is the $V$ band flux.  

\citet{hac25kw} analyzed the $B$, $V$, and $I_{\rm C}$ light curves of
KT Eri in detail and calculated $V$ model light curves. 
They concluded that the first and second
plateaus are the manifestation of the bright accretion disk 
with either irradiation or viscous heating, respectively. 
Each part is indicated by ``irradiated disk''
(magenta line) or ``viscous heating'' (magenta+yellow line)
in Figure \ref{kt_eri_bvi_light_curves}b.

Figure \ref{kt_eri_bvi_light_curves}b shows the interpretation on
the $V$ light curve of KT Eri.  The model light curves
are calculated by \citet{hac25kw}.
They obtained a composite light curve model for a $1.3 ~M_\sun$ WD, that is, 
a combination of free-free emission from the nova winds,
photospheric luminosities of the WD, 
and irradiated accretion disk and companion star, that is,
\begin{eqnarray}
L_{V, \rm total} &=& L_{V, \rm ff,wind} + L_{V, \rm ph, WD} \cr
  & & + L_{V, \rm ph, disk} + L_{V, \rm ph, comp},
\label{luminosity_summation_wd_disk_comp_v-band}
\end{eqnarray}
where $L_{V, \rm ph, WD}$ is the $V$ band flux from the WD photosphere,
$L_{V, \rm ph, disk}$ is the $V$ flux from the optically-thick disk
surface (photosphere), $L_{V, \rm ph, comp}$ the $V$ flux from
the companion star photosphere, and  $L_{V, \rm ff,wind}$ is 
the $V$ flux from free-free emission outside the WD photosphere, 
which is approximately calculated by
\begin{equation}
L_{V, \rm ff,wind} = A_{\rm ff} ~{{\dot M^2_{\rm wind}}
\over{v^2_{\rm ph} R_{\rm ph}}}.
\label{free-free_flux_v-band}
\end{equation}
This $L_{V, \rm ff,wind}$ represents the flux of free-free emission
from optically thin plasma just outside the photosphere,
and $\dot{M}_{\rm wind}$ is the wind mass-loss rate,
$v_{\rm ph}$ the velocity at the photosphere,
and $R_{\rm ph}$ the photospheric radius of the WD. 
The optically-thick nova winds are accelerated deep inside the photosphere,
but the winds become optically thin outside the photosphere.
See \citet{hac20skhs} for the derivation of this formula
and the coefficient $A_{\rm ff}$.

The black line labeled free-free denotes the WD components of
$L_{V, \rm ff,wind} + L_{V, \rm ph, WD}$. 
In the early phase, the WD photosphere expands to a giant star size
and engulfs the companion star.  Strong winds blow.
Free-free emission from the winds dominates the nova spectrum.
Then the WD photosphere gradually shrinks.
The irradiation effect of the disk becomes the main optical source
after the disk emerges from the WD photosphere,
as shown by the magenta line in Figure \ref{kt_eri_bvi_light_curves}b.  
Figure \ref{kt_eri_config} shows our binary configuration after
the disk emerges from the WD photosphere.

Around slightly before the optically thick winds ended on day 103,
the nova enters the supersoft X-ray source (SSS) phase.
There are no bright optical sources except the irradiated accretion disk
and companion star, that is, $L_{V, \rm ph, disk} + L_{V, \rm ph, comp}$.
The model supersoft X-ray light curve is depicted by the thin magenta line
in Figure \ref{kt_eri_bvi_light_curves}b.  The X-ray flux (0.3--10.0 keV)
is calculated with the blackbody approximation of the WD photosphere.

%\placefigure
%Fig.4 
\begin{figure*}
\epsscale{0.8}
%%\plotone{hr_diagram_ASASSN16oh_vi_millinovae_outburst.epsi}
\plotone{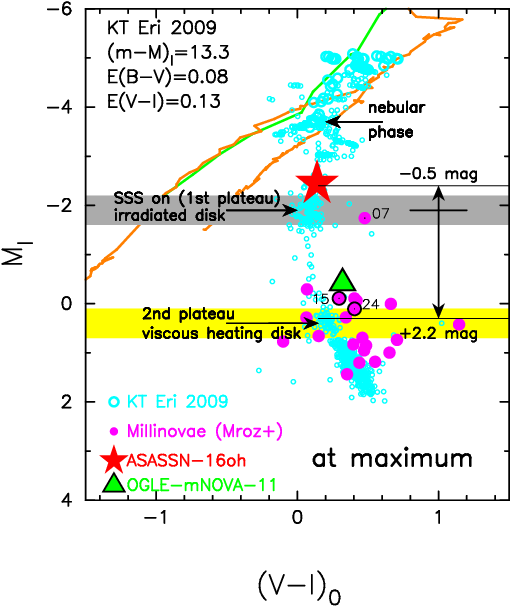}
\caption{
The $(V-I)_0$-$M_I$ color-magnitude diagram track of KT Eri
(open cyan circles) and 23 millinova candidates
(among the total 29 millinova candidates)
at the outburst peak \citep[filled magenta circles; ][]{mro24ks}.
Various time epochs of outburst evolution are shown for KT Eri,
but only each peak position is depicted for the 23 millinovae \citep{mro24ks}.
The filled red star is the peak position of ASASSN-16oh while the filled
green triangle is the peak of OGLE-mNOVA-11, both of which are detected
in X-ray ($\gtrsim 1\times 10^{35}$ erg s$^{-1}$).  
The millinova labeled ``07'' (encircled ``15'' and ``24'') corresponds to
OGLE-mNOVA-07 (-15 and -24) in Table 1 of \citet{mro24ks}.
The two tracks of V1500 Cyg (thick green line) and LV Vul
(thick orange line) are added, taken from Figure 12(b) of \citet{hac25kw}.
The horizontal broad gray belt shows the brightnesses in the first
plateau (SSS) phase while the yellow belt corresponds to the brightnesses
in the second plateau (viscous heating) phase of KT Eri.
The effect of inclination angle of binaries are added by the two arrows
labeled ``$-0.5$ mag'' and ``$+2.2$ mag.''
%The vertical solid red line of $(V-I)_0= +0.22$ is the
%intrinsic color of optically thick free-free emission \citep{hac21k}.
See the main text for more details.
\label{hr_diagram_ASASSN16oh_vi_millinovae_outburst}}
\end{figure*}

\subsection{Comparison of ASASSN-16oh and KT Eri}

Although KT Eri is a classical nova while ASASSN-16oh is a dwarf nova,
these two objects show close resemblances.
The WD mass of ASASSN-16oh is estimated by \citet{kat20sh} to be
1.32 $M_\sun$, whereas $1.3~M_\sun$ in KT Eri \citep{hac25kw}. 
The orbital period of KT Eri is 2.6 days \citep{schaefer22wh} 
and \citet{raj17cb} reported 5.66 days for ASASSN-16oh, as mentioned
in Section \ref{sec_introduction}. 

Figure \ref{as16oh_vi_linear} compares
the $I$ light curve of ASASSN-16oh with the $I_{\rm C}$ light curve of 
KT Eri at the same distance modulus of SMC.  Here  
we assume that the distance moduli in the $I$ band
of ASASSN-16oh and KT Eri are $(m-M)_I=19.0$ and 13.3, respectively, 
i.e., KT Eri is placed 5.7 mag down in the figure. 
We also shift the time of KT Eri in the horizontal direction 
to fit its decay of soft X-ray light curve with
that of ASASSN-16oh. 
In this figure, we plot only the middle and late phases of KT Eri
because we selectively show the first and second plateau phases of KT Eri. 

From the comparison in Figure \ref{as16oh_vi_linear}, 
we are able to list the following points: 
\begin{itemize}
\item[1.] The duration of the SSS phase of KT Eri is in good 
agreement with the outburst width of ASASSN-16oh.
\item[2.] The brightness of the first plateau of KT Eri is close to,
but slightly below, the peak brightness of ASASSN-16oh.
\item[3.] The brightness in the second plateau of KT Eri agrees with 
  the brightness in the post-outburst of ASASSN-16oh.
\item[4.] The $I$ magnitude decline of KT Eri from the first plateau
to the second plateau is well overlapped to that of ASASSN-16oh.
This suggests that the cooling timescales both of the WDs
are almost the same after hydrogen burning ends.
\item[5.] The brightness in quiescence of KT Eri ($I_{\rm C}\sim 14.54
\rightarrow ~M_I\sim 1.24$)
  broadly agrees with   the brightness in quiescence of ASASSN-16oh
  ($I\sim 20.27 \rightarrow ~M_I\sim 1.27$).
\end{itemize}
These provide important clues to understand the nature of millinovae. 

Both the first and second plateaus in KT Eri are explained with/without
irradiation by the central WD \citep{hac25kw}. 
The peak brightness of ASASSN-16oh, similar to
the KT Eri first plateau, is consistent with that (1) hydrogen burning 
occurs on the WD and (2) the irradiated disk has a disk surface area 
similar to KT Eri. It is supported from the suggestion that 
both the binary systems have similar large orbital periods 
(a few to several days).

The brightness of the second plateau in the KT Eri is determined by the
mass-accretion rate through the disk to the WD because 
hydrogen burning already ended.  
We expect a similar accretion rate 
to KT Eri in the post-outburst phase of ASASSN-16oh, 
during at least  300 days, from HJD 2,457,800 to HJD 2,458,100. 

Similar cooling timescales of $I$ light curves in KT Eri and ASASSN-16oh
indicate the similar (almost the same) WD masses, 
because the cooling timescale depends strongly
on the WD mass.  \citet{hac25kw} estimated the WD mass to be 
$1.3 ~M_\sun$ for KT Eri while \citet{kat20sh} suggested $1.32 ~M_\sun$
for ASASSN-16oh. 
These two WD mass estimates are consistent with each other.
Thus, if ASASSN-16oh hosts a hydrogen burning WD, we may conclude that  
nuclear burning extinguished soon after the optical peak.  Then, the
decay from the peak to post-outburst in the $V/I$ light curves
indicates the weakening irradiation effect by the cooling WD.

\section{Demarcation criterion for hydrogen burning}
\label{demarcation_hydrogen_burning}

\subsection{Peak $I$ brightnesses of millinovae}
\label{peak_brightnesses_millinovae}

ASASSN-16oh shows a large outburst amplitude of 
$\Delta I \equiv I_{\rm q} - I_{\rm max} = 3.7 $ mag, where
$I_{\rm q}$ is the mean $I$ magnitude in quiescence and
$I_{\rm max}$ the peak $I$ magnitude of the outburst
\citep[see Table 1 of ][]{mro24ks}. 
Because the orbital period of ASASSN-16oh is as long as 5.66 days
\citep{raj17cb}, its companion star has already evolved to 
a subgiant and its mass transfer rate could be much larger than
those in typical dwarf novae.
ASASSN-16oh could host a large accretion disk around the WD. 
Thus, dwarf nova outbursts of ASASSN-16oh could be much brighter
than typical dwarf novae if the disk is irradiated by the hydrogen
burning hot WD.

\citet{mro24ks} found 29 millinovae candidates and 
summarized various physical quantities including the peak $I_{\rm max}$
brightness and its color $(V-I)_{\rm max}$. 
We plot their 23 millinovae (among the total 29 millinovae)
in the $(V-I)_0$--$M_I$ color-magnitude diagram in 
Figure \ref{hr_diagram_ASASSN16oh_vi_millinovae_outburst} 
together with the evolution track of KT Eri.  
ASASSN-16oh is indicated by the large filled red star mark.

\begin{figure*}
\epsscale{0.8}
%%\plotone{hr_diagram_ASASSN16oh_vi_millinovae_outburst.epsi}
\plotone{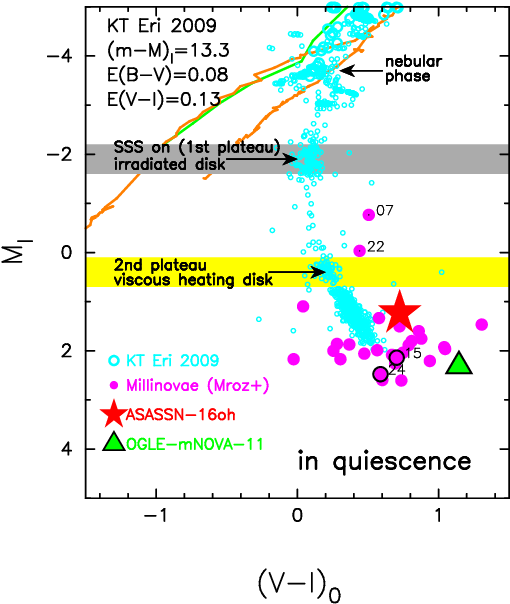}
\caption{
Same as Figure \ref{hr_diagram_ASASSN16oh_vi_millinovae_outburst},
but for quiescent phases of the 29 millinova candidates.
The two millinovae labeled ``07'' and ``22'' corresponds to
OGLE-mNOVA-07 and OGLE-mNOVA-22 \citep{mro24ks}, respectively.
The two millinovae of 07 and 22 are brighter than the yellow belt
(the second plateau of KT Eri) and could not be the members of the SMC or LMC.
% 07 :: -0.76 ==> -1.74 mag  22:: -0.03 ==> -0.95
\label{hr_diagram_ASASSN16oh_vi_millinovae_quiescent}}
\end{figure*}

KT Eri evolves downward and stays at the first plateau of 
$M_I \sim -1.9\pm0.3$ during the SSS phase.  Here, $\pm 0.3$ mag
represents scatters in the $I_{\rm C}$ magnitude distribution during
the first plateau (see Figures 
\ref{hr_diagram_ASASSN16oh_vi_millinovae_outburst} and
\ref{hr_diagram_ASASSN16oh_vi_millinovae_quiescent}).
Its brightness can be reproduced by an accretion disk irradiated by the
hydrogen burning hot WD, as shown in Figure \ref{kt_eri_bvi_light_curves}b
with a model of Figure \ref{kt_eri_config}b.
The peak $I$ magnitude of ASASSN-16oh is located slightly above 
the first plateau of KT Eri, as in Figure 
\ref{hr_diagram_ASASSN16oh_vi_millinovae_outburst}. 
We therefore may conclude that, if ASASSN-16oh hosts a hydrogen-burning
WD, its peak $I$ brightness can be explained by an accretion disk irradiated 
by this hot WD. 

Thus, we define the demarcation line for hydrogen-burning
to be $M_{I,1}= -1.9\pm 0.3$ (horizontal gray belt zone).

Next, we set another demarcation line, 
$M_{I,2}= 0.3\pm 0.3$ (horizontal yellow belt zone). 
This corresponds to the $I$ brightness of the second plateau of KT Eri. 
Because the hydrogen burning has already ended,
the $I$ brightness in the second plateau can be explained by viscous heating
in the accretion disk.  The brightness depends mainly on the mass-accretion
rate.  \citet{hac25kw} reproduced the $V$ brightness in the second plateau
of KT Eri by assuming the mass-accretion rate of
$\dot{M}_{\rm acc}= 2\times 10^{-7} ~M_\sun$ yr$^{-1}$. 
It should be noted again that this rate is close to
the lower bound for steady hydrogen burning, and could give 
an upper bound for the brightness of viscous heating.  
If the mass accretion rate is larger than 
$\dot{M}_{\rm acc}\approx 2.4\times 10^{-7} ~M_\sun$ yr$^{-1}$,
hydrogen burns steadily (as shown in Figure \ref{kt_eri_bvi_light_curves}b)
and the $V$ brightness keeps $M_{V,1}\sim -1.8$,
much brighter than the $V$ brightness in the second plateau,
$M_{V,2}\sim +0.5$.

In Figure \ref{hr_diagram_ASASSN16oh_vi_millinovae_outburst},
many millinova candidates are located $\sim 2$ mag below the position of
ASASSN-16oh, and close to or on the yellow belt line.  Moreover, 
its distribution is scattered with a width of more than 2 mag. 
This suggests that no hydrogen burning occurs in some fainter objects 
and some candidates brighten up only by viscous heating.

%\placefigure
%Fig.5 

\subsection{Dependence on the binary inclination}
\label{depend_binary_inclination}

The brightness of the disk depends on the inclination angle $i$ of
a binary \citep[e.g., $i= 41.2\arcdeg$ for KT Eri: ][]{hac25kw}.    
For dwarf novae at or near maximum light, the surface of the accretion disk
can be assumed to be optically thick, and in this case, \citet{pac80s}
found that, for a binary with the inclination angle of $i$,
the magnitude of a flat, limb-darkened
(with $u =  0.6$) disk should be corrected by
\begin{eqnarray}
\Delta M_I(i) = &-&2.5\log [(1 + {3\over 2} \cos i) \cos i ] \cr
&+&2.5\log [(1 + {3\over 2} \cos 41.2\arcdeg) \cos 41.2\arcdeg ].
\label{effect_inclination}
\end{eqnarray}
This shows a strong dependence on the inclination angle:
a pole-on disk is about 0.5 mag brighter than KT Eri,
while stars with deep eclipses (typically $ i > 80\arcdeg$) 
should be at least 2.2 mag fainter than that of KT Eri.

The demarcation line moves up and down if we include the effect of the
inclination, especially, toward a high inclination binary.  We suppose
that supersoft X-rays are originated from a hydrogen burning WD.  
If a soft X-ray flux of $L_X \gtrsim 1\times 10^{35}$ erg s$^{-1}$
is detected from a millinova, it indicates that hydrogen burns
on the surface of the WD even if its optical peak is $\sim 2$ mag fainter
than the gray belt demarcation line (first plateau of KT Eri), as shown
in Figure \ref{hr_diagram_ASASSN16oh_vi_millinovae_outburst}.
For a low inclination angle (view from near pole-on), on the other hand,
its optical peak could be $\sim 0.5$ mag brighter than the gray belt
demarcation line.  This covers the position of ASASSN-16oh. 

To summarize, for a transient supersoft X-ray source (millinova) with 
$L_X\gtrsim 1\times 10^{35}$ erg s$^{-1}$ or so, we regard that hydrogen burns
on the surface of the WD when it is brighter than $M_{I,1, \rm faint}\equiv 
M_{\rm I,1}+2.2 = -1.9 + 2.2 =0.3$ mag.
If no X-rays are detected and fainter than 
$M_{I,\rm peak} \gtrsim M_{I,2} = +0.3$, 
we suppose that this object belongs to normal dwarf novae.
The large scatter of millinova $M_{I,\rm peak}$
in Figure \ref{hr_diagram_ASASSN16oh_vi_millinovae_outburst}
may reflect the scatter in the inclination angle.

If our interpretation is correct, there is a trend that
the fainter the $M_{I,\rm peak}$,
the smaller the supersoft X-ray flux $L_X$. 
This is because the photosphere of the WD is more shielded by the
disk edge for a higher inclination binary.

It is interesting that the peak position of OGLE-mNOVA-11
(filled green triangle) is upper than $M_{I,1, \rm faint}= +0.3$ mag (thin
horizontal black line labeled ``+2.2 mag'' 
in Figure \ref{hr_diagram_ASASSN16oh_vi_millinovae_outburst}).
An X-ray flux of $L_X\sim 1\times 10^{35}$ erg s$^{-1}$ is detected.
We may conclude that hydrogen burns on the WD but a large part of
the WD photosphere is occulted by a flaring-up disk edge or something else 
\citep[see, e.g., discussion of ][]{hil19}.

\subsection{Quiescent brightnesses of millinovae}
\label{quiescent_brightnesses_millinovae}

Figure \ref{hr_diagram_ASASSN16oh_vi_millinovae_quiescent} shows 
the positions of 29 millinovae candidates in quiescence \citep{mro24ks}.
Many of them are located below the second plateau (yellow belt line)
in the $(V-I)_0$-$M_I$ color-magnitude diagram and their brightnesses
are consistent with the quiescent brightness of KT Eri.  
This indicates that, in quiescence, the mass-accretion rate is
substantially lower than 
$\dot{M}_{\rm acc}=2\times 10^{-7} ~M_\sun$ yr$^{-1}$
unless the inclination angle is as high as $i \gtrsim 75\arcdeg$.

There are two millinovae above the second plateau of KT Eri.
These two millinovae labeled ``07'' and ``22'' correspond to
OGLE-mNOVA-07 and OGLE-mNOVA-22 \citep{mro24ks}, respectively.
If these two millinovae shine by viscous heating in the accretion disk,
their mass-accretion rates should be much higher than 
$\dot{M}_{\rm acc}=2\times 10^{-7} ~M_\sun$ yr$^{-1}$.
With such a high mass-accretion rate, hydrogen burning is stable
as shown in Figure \ref{kt_eri_bvi_light_curves}b, and
the disk should be irradiated all the time.  In this case, we expect
a constant brightness of $M_{I,1}= -1.9$ (or $M_{V,1}= -1.8$, 
yellow+orange line in Figure \ref{kt_eri_bvi_light_curves}b).
Therefore, we may conclude that these two millinovae candidates are not
millinovae, but much fainter dwarf novae in our Galaxy (foreground stars).
The rise in the brightness from quiescence to maximum
is $\Delta I \equiv I_{\rm q} - I_{\rm max}= 0.98$ in OGLE-mNOVA-07
and $\Delta I \equiv I_{\rm q} - I_{\rm max}= 0.92$ in OGLE-mNOVA-22,  
% 07 :: -0.76 ==> -1.74 mag  22:: -0.03 ==> -0.95
% 01 :: 20.27 ==> 16.60 mag  11:: 20.88 ==> 18.15
being much smaller than the other millinovae, for example,
$\Delta I = I_{\rm q} - I_{\rm max}= 3.68$ in ASASSN-16oh (OGLE-mNOVA-01) and
$\Delta I = I_{\rm q} - I_{\rm max}= 2.73$ in OGLE-mNOVA-11.
Thus, the yellow belt demarcation line can be utilized to distinguish
millinovae from typical foreground dwarf novae.

In this way, we utilize our demarcation line of the second plateau
(viscous heating) to distinguish millinovae in the SMC and LMC from
foreground dwarf novae in our Galaxy.

\subsection{Amplitudes of the millinova outbursts}
\label{amplitude_millinova_outburst}

Millinovae stay below the second plateau line of $M_{I,2}= +0.3$ because
the mass accretion rate is smaller than $\dot{M}_{\rm acc}\sim
2\times 10^{-7} ~M_\sun$ yr$^{-1}$.  If they go into outburst by triggering
thermal instability of a disk and hydrogen burns on the WDs,
their brightnesses reach at least $M_{I,2}= -1.9$ like in KT Eri
because of irradiation effect.
Thus, their amplitudes of outburst should exceed
\begin{equation}
\Delta I \equiv I_{\rm q} - I_{\rm max} \gtrsim M_{I,2} - M_{I,1} = 2.2.
\label{amplitude_I_mag}
\end{equation}
%where $I_{\rm q}$ is the $I$ magnitude in quiescence and
%$I_{\rm max}$ the $I$ magnitude at maximum.
The amplitude of ASASSN-16oh, $\Delta I = I_{\rm q} - I_{\rm max}= 3.7$, 
satisfies this requirement \citep[see $\Delta I$ in Table 1 of ][]{mro24ks}.
OGLE-mNOVA-11 also has the amplitude of 
$\Delta I= I_{\rm q} - I_{\rm max}= 2.7$
and the X-ray fluxes of $L_X\sim 1\times 10^{35}$ erg s$^{-1}$ were detected.
We conclude that hydrogen burns on the WD for mNOVA-11.
Two millinovae candidates show marginal amplitudes of 
$\Delta I = I_{\rm q} - I_{\rm max}= 2.2$ (mNOVA-15) and 2.4 (mNOVA-24).
If an X-ray flux of $L_X\gtrsim 1\times 10^{35}$ erg s$^{-1}$ 
is observed, we conclude that hydrogen burns in these
millinovae (encircled magenta symbols labeled ``15'' and ``24'' in
Figures \ref{hr_diagram_ASASSN16oh_vi_millinovae_outburst} and
\ref{hr_diagram_ASASSN16oh_vi_millinovae_quiescent}). 

A millinova candidate of $\Delta I = I_{\rm q} - I_{\rm max}\lesssim 2$ 
is unlikely to host a hydrogen burning WD.  We do not expect
bright X-ray fluxes of $L_X \gtrsim 1\times 10^{35}$ erg s$^{-1}$.
Therefore, it is not a millinova by definition. 

In this way, the outburst amplitude can be used to 
find a millinova.  If the amplitude of an outburst satisfies 
$\Delta I = I_{\rm q} - I_{\rm max}\gtrsim 2.2$
(Equation (\ref{amplitude_I_mag})) and the peak $I$ magnitude is 
brighter than $M_{I,1,\rm faint}= +0.3$ mag,
it is possibly a millinova 
and an X-ray flux of $L_X\gtrsim 1\times 10^{35}$ erg s$^{-1}$ 
would be detected.  Note that the requirement of Equation
(\ref{amplitude_I_mag}) is independent of the inclination angle $i$
of the binary.

\section{Discussion}
\label{discussion}

\subsection{A Dwarf Nova Triggers a Nova Outburst?}
\label{trigger_nova_outburst}

\citet{kat20sh} discussed the possibility that a dwarf nova outburst
can trigger a nova outburst.  They showed that an accreting WD 
experiences a nova outburst when the mass of the hydrogen-rich envelope
reaches a critical value \citep[see, e.g., Table 1 of ][]{kat20sh}. 
%Table \ref{table_models} list the amount of accreted matter 
%to trigger hydrogen ignition.
The critical masses are a few to several times $10^{-7} ~M_\sun$
for the case of ASASSN-16oh, that is, for a $1.32 ~M_\sun$ WD with
the heavy element content of $Z=0.001$
(or $1.35 ~M_\sun$ WD and $Z=0.0001$).  This low metallicity corresponds to
a typical metallicity of the SMC ([Fe/H]=$-1.25 \pm 0.01$) \citep{chi09}. 
In general, the ignition mass is smaller for a larger mass-accretion
rate of $\dot{M}_{\rm acc}$ (or a shorter recurrence period of $t_{\rm rec}$) 
because the WD is hotter. 

It is useful to see how the hydrogen-rich envelope decreases its mass  
with time in KT Eri. The envelope mass was 
$M_{\rm env,max}= 2.9\times 10^{-6} ~M_\sun$ at optical maximum,
and decreases to 
$M_{\rm env,ws}= 3.2\times 10^{-7} ~M_\sun$ when winds stop. 
Thus, for a hydrogen shell flash on a similar WD mass, wind mass loss
occurs when the ignition mass is larger than 
$M_{\rm env,ws}= 3.2\times 10^{-7} ~M_\sun$. 
The envelope mass in KT Eri further decreases to
$M_{\rm env,hb}= 1.9\times 10^{-7} ~M_\sun$
when the hydrogen burning stops.
This means that, if the ignition mass is between 
$3.2\times 10^{-7} ~M_\sun$ and $1.9\times 10^{-7} ~M_\sun$, 
the nuclear burning ignites on the WD with no mass ejection. 
This broadly fit with the definition of millinova. 
If the hydrogen-rich envelope mass is smaller than 
$1.9\times 10^{-7} ~M_\sun$, hydrogen never ignites. 

Thus, we can divide a thermonuclear runaway event by the amount of
the ignition mass into a classical nova, millinova, and dwarf nova
(no runaway). 
The corresponding brightnesses are roughly indicated with the upward arrows
labeled ``classical novae'', ``millinovae'', and ``dwarf novae''
in Figure \ref{kt_eri_bvi_light_curves}b. 

The mass-accretion rate onto the WD is estimated in several dwarf novae
by \citet{kim18}.  Among them, V364~Lib shows a relatively large mass
inflow rate of $1 \times 10^{-7}~M_\sun$ yr$^{-1}$.  
This object has an orbital period of $P_{\rm orb}= 0.70$ days, 
rising (decay) time of 10 (35) days, and small outburst amplitude of 1 mag.
If we apply this mass-accretion rate to KT Eri, the total accreted mass
reaches $((10+35)/365)\times (1 \times 10^{-7}~M_\sun$ yr$^{-1})= 1.2\times
10^{-8} ~M_\sun$ at most during the outburst,
being too small to ignite hydrogen on the WD.

In the case of ASASSN-16oh, we expect a much larger mass inflow rate 
as suggested from a much larger outburst amplitude
($\Delta I = I_{\rm q}-I_{\rm max}= 3.68$) and longer orbital period
\citep[5.66 days,][]{raj17cb}.  If we assume the 
mass inflow rate of several times $10^{-7}~M_\sun$ yr$^{-1}$, the WD may
accrete sufficient mass for hydrogen ignition, that is, $(200/365)\times
(5 \times 10^{-7}~M_\sun$ yr$^{-1})=2.7\times 10^{-7}~M_\sun$
before maximum (see Figure \ref{as16oh_vi_linear}).
This is between $3.2\times 10^{-7} ~M_\sun$ and $1.9\times 10^{-7} ~M_\sun$, 
and satisfies the condition of millinova.

\subsection{The brightness of the first plateau}
\label{brightness_first_plateau}

The brightnesses of the first and second plateau, $M_{I,1}$ and 
$M_{I,2}$ could depend not only on the inclination angle $i$ but
also on the other binary parameters.  In this subsection, we discuss
the dependence of the first plateau brightness ($M_{V,1}$ or $M_{I,1}$)
on the WD mass, metallicity, and orbital period. 

\subsubsection{white dwarf mass and metallicity}
\label{wd_mass_metallicity}

The WD luminosity in the SSS phase is estimated to be close to the
Eddington luminosity \citep[e.g.,][]{kat22shb},
\begin{equation}
L_{\rm ph} \approx L_{\rm Edd}\equiv {{4\pi c G M_{\rm WD}}
\over \kappa_{\rm el}},
\label{eddington_luminosity}
\end{equation}
where $L_{\rm ph}$ is the photospheric luminosity of the WD, $c$ the speed
of light, $G$ the gravitational constant, $M_{\rm WD}$ the mass of the WD,
and $\kappa_{\rm el}$ the electron scattering opacity.
The irradiation effect is mainly determined by the luminosity of the
central WD and therefore the brightness of $M_{V,1}$ or $M_{I,1}$ depends
approximately on $M_{\rm WD}/\kappa_{\rm el}$.  In the SSS phase,
the photospheric temperature of the WD is as high as $\sim 10^6$ K and
hydrogen and helium atoms are completely ionized at the photosphere.
Then, the opacity is expressed by $\kappa_{\rm ph}\approx \kappa_{\rm el}=
0.2(1+X)$ cm$^{-2}$ g$^{-1}$, where $\kappa_{\rm ph}$ is the opacity at 
the photosphere and $X$ is the hydrogen content (by mass weight).
Thus, the metallicity difference between LMC/SMC and our Galaxy
does not make a significant difference in the brightness of
$M_{V,1}$ or $M_{I,1}$.  

The WD mass can be restricted from the condition that 
mass accretion by single dwarf nova outburst triggers 
nuclear burning on the WD. 
We estimate the amount of mass supplied by the dwarf nova instability 
in ASASSN-16oh to be as small as several times $10^{-7} ~M_\sun$
(see Section \ref{trigger_nova_outburst}). 
This value may correspond to the largest accreted mass for millinovae 
because ASASSN-16oh is the brightest one among the 29 millinovae 
candidates \citep{mro24ks}. 

The ignition mass is defined by the mass of a hydrogen-rich envelope
when thermonuclear runaway starts on a WD, and is presented for various
WD masses and mass accretion rates in Figure 2 of \citet{kat21sh}. 
The ignition mass depends on the mass accretion rate, but 
the ignition mass of several times $10^{-7} ~M_\sun$ appears only on
the WD mass of $M_{\rm WD} > 1.3 ~M_\sun$. 
A less massive WD ($M_{\rm WD} < 1.3 ~M_\sun$) needs
repeated dwarf nova outbursts 
before the envelope mass reaches the required ignition mass. 
In this case, however, hydrogen burning lasts much longer
than the case of ASASSN-16oh
\citep[e.g., for a $1.2 ~M_\sun$ WD, its duration is 10 times 
longer, see ][] {hil19,kat20sh}. 
Thus, we can exclude WD masses of  $M_{\rm WD}< 1.3 ~M_\sun$
if a millinova hosts a hydrogen burning WD.

The brightness of the first plateau depends linearly
on the WD mass (see Equation (\ref{eddington_luminosity})).
Because a hydrogen burning millinova hosts a very massive WD, 
the brightness difference by the WD mass is smaller than
5\%, that is, 0.1 mag in the $M_{V,1}$ or $M_{I,1}$.

%Table 1
%\placetable{table_brightness_first_plateau}

%%%\startlongtable
\begin{deluxetable}{lrrrr}
\tabletypesize{\scriptsize}
\tablecaption{Brightnesses of the first and second plateau
phases\tablenotemark{a}
\label{table_brightness_first_plateau}}
\tablewidth{0pt}
\tablehead{
\colhead{$P_{\rm orb}$} 
%%%& \colhead{$M_{\rm WD}$} & \colhead{$M_2$}
%%%& \colhead{$T_{\rm ph,2}$} & \colhead{$\dot{M}_{\rm acc}$}
& \colhead{$M_{V,1}$} & \colhead{$M_{V,2}$} 
& \colhead{$M_{I,1}$} & \colhead{$M_{I,2}$} \\
(day) & (mag) & (mag) & (mag) & (mag)
}
\startdata
5.66 & $-2.4$ & 0.0 & $-2.5$ & $-0.2$ \\
2.62 & $-1.8$ & 0.5 & $-1.9$ & 0.3 \\
1.23 & $-1.1$ & 1.2 & $-1.2$ & 1.0
\enddata
\tablenotetext{a}{
$M_{\rm WD}=1.3 ~M_\sun$; $M_2=1.0 ~M_\sun$; $T_{\rm ph,2}=4500$ K;
the inclination angle of the binary $i=41.2\arcdeg$;
$\dot{M}_{\rm acc}=2\times 10^{-7} ~M_\sun$ yr$^{-1}$;
$P_{\rm orb}$ is the orbital period of the binary; 
$M_{V,1}$/$M_{I,1}$ the absolute $V/I$ brightnesses at the first plateau;
$M_{V,2}$/$M_{I,2}$ the absolute $V/I$ brightnesses at the second plateau.
}
%\tablenotetext{b}{
%Orange line in Figure \ref{v392_per_only_v_x_big_disk_6100k_logscale}b.
%}
%\tablenotetext{c}{
%The thick orange line in Figure \ref{kt_eri_v339_del_v_logscale_no3_ab}a.
%}
%\tablenotetext{d}{
%The thin red line in Figure \ref{kt_eri_v339_del_v_logscale_no3_ab}a.
%}
%\tablenotetext{e}{
%The magenta line in Figure \ref{kt_eri_v339_del_v_logscale_no3_ab}b.
%}
%\tablenotetext{f}{
%The dark gray line in Figure \ref{kt_eri_v339_del_v_logscale_no3_ab}b.
%}
%\tablenotetext{g}{
%The thick orange line in Figure \ref{kt_eri_v339_del_v_logscale_no3_ab}b.
%}
%\tablenotetext{h}{
%The thin red line in Figure \ref{kt_eri_v339_del_v_logscale_no3_ab}b.
%}
%\tablenotetext{i}{
%The magenta line in Figure \ref{kt_eri_only_v_x_big_disk_4500k_logscale}b.
%}
%\tablenotetext{j}{
%The orange line in Figure \ref{kt_eri_only_v_x_big_disk_4500k_logscale}b.
%}
%\tablenotetext{k}{
%The orange line in Figure \ref{kt_eri_only_v_x_big_disk_4500k_logscale}b.
%}
\end{deluxetable}

\subsubsection{orbital period}
\label{binary_orbital_period}

The brightnesses at the first and second plateaus depend on the orbital
period $P_{\rm orb}$ of the binary because a large $P_{\rm orb}$ binary
could have a large accretion disk.
We have calculated the $V$ light curves using the KT Eri model but for
two different orbital periods of 5.66 days (ASASSN-16oh) and 1.23 days
(U Sco).  Table \ref{table_brightness_first_plateau} shows the 
brightnesses of $M_{V,1}$, $M_{V,2}$, $M_{I,1}$, and $M_{I,2}$
for these orbital periods including the $P_{\rm orb}=2.62$ days of KT Eri.
The other binary parameters are assumed to be the same
as those in KT Eri.  For $P_{\rm orb}= 5.66$ days,
$\Delta M_{V,1} = -0.6$ mag brighter, but 
for $P_{\rm orb}= 1.23$ days,
$\Delta M_{V,1} = +0.7$ mag fainter than in KT Eri.

It is important that the brightness difference between the first plateau
and the second plateau is still $M_{V,2} - M_{V,1} = 2.2$ or 2.3 and
$M_{I,2} - M_{I,1} = 2.2$ or 2.3 among the three $P_{\rm orb}$ cases.
Therefore, our requirement of $I_{\rm q} - I_{\rm max} \gtrsim 2.2$ mag
(Equation (\ref{amplitude_I_mag}) in 
Section \ref{amplitude_millinova_outburst}) is safely satisfied.
The peak $I$ brightness of ASASSN-16oh is 
$M_{I,\rm max}= -2.4$ (the inclination angle $i$ is not known)
and consistent with our result of $M_{I,1}= -2.5$ ($i=41.2\arcdeg$
in KT Eri) in Table \ref{table_brightness_first_plateau}.    

\citet{mro24ks} suggested the decay timescales ($\tau_{\rm d}$) of millinovae
are between $20 \lesssim \tau_{\rm d} \lesssim 70$ days mag$^{-1}$ and
this range corresponds roughly to $3 \lesssim P_{\rm orb} \lesssim 16$ days
from the relation in their Figure 6.  Only two millinovae have known 
orbital periods of 5.66 days (ASASSN-16oh) and 4.83 days (OGLE-mNOVA-08).
We suppose that ASASSN-16oh has the longest orbital period because it is
the brightest one among the 29 millinova candidates.
On the other hand, KT Eri has the orbital period of 2.62 days,
which corresponds possibly to the shortest case.
Therefore, the range of $M_{I,1}$ is between
$-2.5$ and $-1.9$ for typical millinovae if they have a hydrogen burning WD.

\subsection{A possible way to Type Ia supernovae?}
\label{possible_way_to_sn1a}

It should be noted that there is an essential difference between
millinovae and classical novae: a millinova has no mass ejection
during the outburst while a classical nova loses almost all accreted
mass by winds.  In other words, the mass retention efficiency is about 100\%
in millinovae.  Because the WD mass of ASASSN-16oh is already as massive as
$1.32 ~M_\sun$, \citet{kat20sh} suggested that ASASSN-16oh is a candidate
of Type Ia supernova progenitors.  When the WD mass reaches $1.38 ~M_\sun$,
it could explode as a Type Ia supernova. 
In this sense, millinovae are also promising objects for Type Ia 
supernova progenitors.

\section{Conclusions}
\label{conclusions}

Our main results are summarized as follows.
\begin{enumerate}
\item We define the demarcation criterion of millinovae for hydrogen burning,
that is, $M_{I,1}= -1.9\pm 0.3$, based on the first plateau of the KT Eri
2009 outburst.
\item ASASSN-16oh satisfies the above demarcation criterion for hydrogen
burning.  Also at optical maximum, its position is just on the KT Eri
track of the 2009 outburst in the $(V-I)_0$-$M_I$
color-magnitude diagram. This confirms that the brightness of ASASSN-16oh
is dominated by an irradiated accretion disk.
\item We also find another demarcation criterion for an upper limit
for viscous heating disks, $M_{I,2}= 0.3\pm 0.3$, based on 
the second plateau of the KT Eri 2009 outburst.
Millinova candidates, the peak magnitudes of which are fainter than
$M_{I,2}$, are highly likely to be a dwarf nova, but not a millinova.
\item There are two millinova candidates substantially above $M_{I,2}$
in quiescence, i.e., OGLE-mNOVA-07 and OGLE-mNOVA-22.
If they were really the members of the LMC, hydrogen should burn steadily
on the WD in quiescence.  Therefore, we may conclude that these
two millinova candidates are not millinovae but much fainter foreground
dwarf novae (in our Galaxy).
In this way, we utilize our second demarcation line of $M_{I,2}$
to distinguish millinovae (nuclear burning object) in the SMC/LMC 
from foreground dwarf novae (viscous heating) in our Galaxy.
\item The brightnesses of demarcation lines depend on the inclination angle
$i$ of the disk (or binary).  The dependency on the angle $i$ is 
given by Equation (\ref{effect_inclination}). 
A pole-on disk is about 0.5 mag brighter than that of
KT Eri ($i=41.2\arcdeg$), while stars with deep eclipses (typically
$ i > 80\arcdeg$) are $\sim 2.2$ mag fainter than KT Eri.
\item To trigger hydrogen burning on a WD by only one dwarf
nova instability, its ignition mass should be as small as $\sim 1\times
10^{-6} ~M_\sun$ or less, which requires the WD mass 
of $M_{\rm WD} \gtrsim 1.3 ~M_\sun$.    
\item Among the 29 millinova candidates, ASASSN-16oh ($P_{\rm orb}=5.66$ days) 
is the brightest at maximum and possibly has the longest orbital period.
On the other hand, KT Eri has the orbital period of 2.62 days,
which corresponds to the shortest one, possibly the faintest one.
Therefore, the range of the first plateau brightness $M_{I,1}$ is between
$-2.5$ and $-1.9$ for typical millinovae if they have a hydrogen burning WD. 
\item When the amplitude of an outburst satisfies $\Delta I \equiv
I_{\rm q} - I_{\rm max}\gtrsim 2.2$ (Equation (\ref{amplitude_I_mag})),
a detection of soft X-ray flux $L_X\gtrsim 1\times 10^{35}$ erg s$^{-1}$ 
supports this argument of an irradiated accretion disk.
The two millinovae, ASASSN-16oh ($\Delta I= 3.7$) and OGLE-mNOVA-11
($\Delta I= 2.7$), satisfy these requirements.
The first requirement of $\Delta I \gtrsim 2.2$
is independent either of the inclination angle or of the orbital 
period of the binary.
We find two more millinovae that marginally satisfy this requirement,
OGLE-mNOVA-15 ($\Delta I= 2.2$) and OGLE-mNOVA-24 ($\Delta I= 2.4$).
\item We predict a tendency; for a fainter $I_{\rm max}$ millinova,
a smaller X-ray flux is observed.
\item For a millinova candidate having a small amplitude of
$\Delta I = I_{\rm q} - I_{\rm max}\lesssim 2$, it is unlikely that
it hosts a hydrogen burning white dwarf (WD).
It is not a millinova by definition.
We do not expect bright X-ray fluxes of 
$L_X \gtrsim 1\times 10^{35}$ erg s$^{-1}$.
\item Because the mass retention efficiency is about 100\%
in millinova outbursts and its WD mass is already as massive as
$\gtrsim 1.3 ~M_\sun$, the WD soon reaches $1.38 ~M_\sun$
and could explode as a Type Ia supernova.  Therefore, 
millinovae are promising candidates for Type Ia supernova progenitors.
\end{enumerate}

\begin{acknowledgments}
We are grateful to the Variable Star Observers League of Japan (VSOLJ) and
the American Association of Variable Star Observers (AAVSO) for the 
archive data of KT Eri.  We also thank the anonymous referee for useful
comments that improved the manuscript.
\end{acknowledgments}

\vspace{5mm}
\facilities{Swift(XRT), AAVSO, SMARTS, VSOLJ, OGLE IV}

%\appendix


\begin{thebibliography}{}

%\bibitem[Abdul-Masih et al. (2019)]{abd19}
%Abdul-Masih, M., Banyard, G., 
%Bodensteiner, J, + et al. 2019, ArXiv:1912.04902
% No signature of the orbital motion of a putative 70 Mo BH in LB-1

%\bibitem[Alcock et al. (1996)]{alc96}
%Alcock, C. et al. 1996, \mnras, 280, L49
%-L53, Optical variability of the Large Magellanic Cloud supersoft
%source RX J0513.9-6951 from MACHO Project photometry

%\bibitem[Bohlin et al. (1978)]{boh78}
%Bohlin, R. C., Savage, B. D., \& Drake, J. F. 1978, \apj, 224, 132
%-142. A survey of interstellar H I from L-alpha absorption measurements. II

% \bibitem[Chandrasekhar (1939)]{cha39} Chandrasekhar, S. 1939, 
% An Introduction to the Study of Stellar Structure", (Chicago:
% University of Chicago Press), or  1957 (New York: Dover).

%\bibitem[Chen et al. (2019)]{che19}
%Chen, H., Woods, T. E., Yungelson, L. R., et al.
%%%%Piersanti, L., Gilfanov, M., Han, Z. 
%2019, \mnras, 490, 1678
%in press (arXiv:1909.17643) 
%Hai-Liang Chen, T. E. Woods, L. R. Yungelson, Luciano Piersanti,
% M. Gilfanov, Zhanwen Han

 
%\bibitem[Chomiuk et al. (2014)]{cho14}
%Chomiuk, L., Nelson, T., Mukai, K., et al. 2014, \apj, 788, 130
%, 13 pp.  The 2011 Outburst of Recurrent Nova T Pyx: 
% X-ray Observations Expose the White Dwarf Mass and Ejection Dynamics

\bibitem[M.-R. L. Cioni (2009)]{chi09}
Cioni, M.-R. L. 2009, \aap, 506, 1137, \doi{10.1051/0004-6361/200912138}
% the metallicity gradient as a tracer of history and structure: LMCs and M33


%\bibitem[Cowley et al. (2002)]{cow02}
%Cowley, A.P., Schmidtke, P.C., Crampton, D., \& Hutchings, J.B. 2002,
%\aj, 124, 2233
%-2237, A new orbital ephemeris and reinterpretation of spectroscopic
% data for the supersoft X-ray binary RX J0513.9-6951

%\bibitem[Darnley et al. (2014)]{dar14}
%Darnley, M. J., Williams, S. C., Bode, M. F., et al. 2014, \aap, 563, L9
%(arXiv:1401.2905)
%Darnley, M. J.; Williams, S. C.; Bode, M. F.; Henze, M.; Ness, J.-U.;
% Shafter, A. W.; Hornoch, K.; Votruba, V.
% A remarkable recurrent nova in M31 - The optical observations

\bibitem[M. J. Darnley et al. (2015)]{dar15}
Darnley, M. J., Henze, M., Steele, I. A., et al. 2015, \aap, 580, A45,
\doi{10.1051/0004-6361/201526027}
% 23 pp. A remarkable recurrent nova in M31: 
% Discovery and optical/UV observations of the predicted 2014 eruption

%\bibitem[Denissenkov et al. (2013)]{den13}
%Denissenkov, P. A., Herwig, F., Bildsten, L., \& Paxton, B. 2013, \apj, 762, 8
% MESA Models of Classical Nova Outbursts: 
% The Multicycle Evolution and Effects of Convective Boundary Mixing

%\bibitem[Diaz (1999)]{dia99}Diaz, M. P. 1999, \pasp, 111, 76
%-83, Time-resolved Spectroscopy of V Sagittae

%\bibitem[Diaz \& Steiner (1995)]{dia95}
%Diaz, M. P., \& Steiner, J. E. 1995, \aj, 110, 1816
%-, The Nova-like Variable WX Centauri and the V Sagittae Phenomenon

%\bibitem[El-Badry \& Quataert (2019)]{elb19}
%El-Badry, K. \& Quataert, E. 2019, ArXiv: 1912.04185
% Not so fast: LB-1 is unlikely to contain 70 Mo BH, submitted to MN

%\bibitem[Ennis et al. (1977)]{enn77}
%Ennis, D., Becklin, E. E., Beckwith, S., et al. 
%%% Elias, J., Gatley, I., Matthews, K., Neugebauer, G., \& Willner, S. P. 
%1977, \apj, 214, 478, \doi{10.1086/155273}
%%-487. Infrared observations of Nova Cygni 1975

\bibitem[P. A. Evans et al. (2009)]{eva09}
Evans, P. A., Beardmore, A. P., Page, K. L., et al.  2009, \mnras, 397, 1177,
\doi{10.1111/j.1365-2966.2009.14913.x}
%-1201. (arXiv: 0812.3662)
%  Methods and results of an automatic analysis
% of a complete sample of Swift-XRT observations of GRBs

%\bibitem[Fujimoto (1982)]{fuj82}
%Fujimoto, M. Y. 1982, \apj, 257, 767
%-350.  A Theory of Hydrogen Shell Flashes on Accreting White Dwarfs 
% - Part Two - the Stable Shell Burning and the Recurrence Period 
% of Shell Flashes

%\bibitem[Gallagher et al. (1980)]{gal80ko}
%%%%\bibitem[Gallagher et al. (1980b)]{gal80b}
%Gallagher, J. S., Kaler, J. B., Olson, E. C., Hartkopf, W. I., \&
%Hunter, D. A. 1980, \pasp,  92, 46, \doi{10.1086/130612}
%%-51. An optical light curve for Nova Cygni 1978

%\bibitem[Gallagher \& Ney (1976)]{gal76}
%Gallagher, J. S., \& Ney, E. P. 1976, \apj, 204, L35, \doi{10.1086/182049}
%%-L39. The early infrared development of Nova Cygni 1975.

%\bibitem[Green et al. (2019)]{gre19}
%Green, G. M., Schlafly, E. F., Zucker, C., et al. 2019, \apj, 887, 93
%\doi{10.3847/1538-4357/ab5362}
%%%arXiv:1801.03555
%% A Three-Dimensional Map of Milky-Way Dust

%\bibitem[Greiner \& van Teeseling (1998)]{gre98}
%Greiner, J., \& van Teeseling, A. 1998, \aap, 339, L21
%-L24, On the X-ray properties of V SGE and its relation
% to the supersoft X-ray binaries

%\bibitem[G\"uver \& \"Ozel (2009)]{guv09}
%G\"uver, T., \& \"Ozel, F. 2009, \mnras, 400, 2050
%-2053. The relation between optical extinction and hydrogen column density
%  in the Galaxy

\bibitem[I. Hachisu \& M. Kato (2001)]{hac01kb}
Hachisu, I., \& Kato, M. 2001, \apj, 558, 323, \doi{10.1086/321601}
%-350.  Recurrent Novae as a Progenitor System of Type Ia Supernovae. I.
%RS Ophiuchi Subclass --- Systems with a Red Giant Companion

%\bibitem[Hachisu \& Kato (2003a)]{hac03RXJ}
%Hachisu, I., \& Kato, M. 2003a, \apj, 590, 445
% RX J0513.9-6951: The First Example of Accretion Wind Evolution,

%\bibitem[Hachisu \& Kato (2003b)]{hac03VSge}
% Hachisu, I., \& Kato, M. 2003b, \apj, 598, 527
% A Limit Cycle Model for Long-Term Optical Variations of V Sagittae:
%    The Second Example of Accretion Wind Evolution,

\bibitem[I. Hachisu \& M. Kato (2006)]{hac06kb}
Hachisu, I., \& Kato, M. 2006, \apjs, 167, 59, \doi{10.1086/508063}
% (Paper I)
%-80. A Universal Decline Law of Classical Novae

%\bibitem[Hachisu \& Kato (2007)]{hac07k}
%Hachisu, I., \& Kato, M. 2007, \apj, 662, 552
%(astro-ph/0702563)
%(Paper II)
%-563.  A Universal Decline Law of Classical Novae.
% II. GK Persei 1901 and Novae in 2005

%\bibitem[Hachisu \& Kato (2010)]{hac10k}
% Hachisu, I., \& Kato, M. 2010, \apj, 709, 680
%  A Prediction Formula of Supersoft X-ray Phase of Classical Novae

%\bibitem[Hachisu \& Kato (2014)]{hac14k}
%Hachisu, I., \& Kato, M. 2014, \apj, 785, 97, \doi{10.1088/0004-637X/785/2/97}
%%%(Paper I)
%%(1-44) The UBV Color Evolution of Classical Novae. I. Nova-giant Sequence
%%in the Color-color Diagram

\bibitem[I. Hachisu \& M. Kato (2015)]{hac15k}
Hachisu, I., \& Kato, M. 2015, \apj, 798, 76, \doi{10.1088/0004-637X/798/2/76}
% (1-29) A Light Curve Analysis of Classical Novae: Free-free Emission vs.
% Photospheric Emission

%\bibitem[Hachisu \& Kato (2016a)]{hac16k}
%Hachisu, I., \& Kato, M. 2016a, \apj, 816, 26,
% \doi{10.3847/0004-637X/816/1/26}
%%%in press (arXiv:1511.06819), pp.69
%% Light Curve Analysis of Neon Novae

%\bibitem[Hachisu \& Kato (2016b)]{hac16kb}
%Hachisu, I., \& Kato, M. 2016b, \apjs, 223, 21, 
%\doi{10.3847/0067-0049/223/2/21}
%%in press (arXiv:1602.01195)
% The UBV Color Evolution of Classical Novae. II. Color-Magnitude Diagram

%\bibitem[Hachisu \& Kato (2017)]{hac17k}
%Hachisu, I., \& Kato, M. 2017, in Proceedings of the Palermo Workshop
%2017 on ``The Golden Age of Cataclysmic Variables and Related
%Objects - IV'', ed. F. Giovannelli et al. (Trieste: SISSA PoS), 315, 47
% https://pos.sissa.it/315/047/pdf
% proceedings of the Palermo Workshop 2017 on ``The Golden Age of
% Cataclysmic Variables and Related Objects - IV''

%\bibitem[Hachisu \& Kato (2018a)]{hac18ka}
%Hachisu, I., \& Kato, M. 2018a, \apj, 858, 108
%16 pp. A Light-curve Analysis of Gamma-Ray Nova V959 Mon:
% Distance and White Dwarf Mass

%\bibitem[Hachisu \& Kato (2018b)]{hac18kb}
%Hachisu, I., \& Kato, M. 2018b, \apjs, 237, 4
%51 pp. A Light Curve Analysis of Recurrent and Very Fast Novae
% in our Galaxy, Magellanic Clouds, and M31

%\bibitem[Hachisu \& Kato (2019a)]{hac19k}
%Hachisu, I., \& Kato, M. 2019a, \apjs, 241, 4, \doi{10.3847/1538-4365/ab0202}
%68 pp. The UBV Color Evolution of Classical Novae. III.
% Time-Stretched Color-Magnitude Diagram of Recent Novae in Outburst
% (Paper III),

%\bibitem[Hachisu \& Kato (2019b)]{hac19kb}
%Hachisu, I., \& Kato, M. 2019b, \apjs, 242, 18
%68 pp. A Light Curve Analysis of 32 Recent Galactic Novae ---
% Distances and White Dwarf Masses

\bibitem[I. Hachisu \& M. Kato (2021)]{hac21k}
Hachisu, I., \& Kato, M. 2021, \apjs, 253, 27, \doi{10.3847/1538-4365/abd31e}
%%%in press (arXiv:2012.06100)
%  The $UBV$ Color Evolution of Classical Novae. IV.
%  Time-Stretched $(U-B)_0$-$(M_B-2.5\log f_{\rm s})$ and
%  $(V-I)_0$-$(M_I-2.5\log f_{\rm s})$ Color-Magnitude Diagrams
%  of Novae in Outburst

\bibitem[I. Hachisu \& M. Kato (2023)]{hac23k}
Hachisu, I., \& Kato, M. 2023, \apj, 953, 78, \doi{10.3847/1538-4357/acdfd3}
%%%in press, arXiv:2306.09218
% A multiwavelength light curve analysis of the classical nova YZ Ret:
% An extension of the universal decline law to the nebular phase

%\bibitem[Hachisu et al. (2008)Hachisu, Kato, \& Cassatella]{hac08kc}
%Hachisu, I., Kato, M., \& Cassatella, A. 2008, \apj, 687, 1236
%(Paper III)
%-1252.  A Universal Decline Law of Classical Novae.
% III. GQ Muscae 1983

\bibitem[I. Hachisu et al. (2000)]{hkkm00}
Hachisu, I., Kato, M., Kato, T., \& Matsumoto, K. 2000, \apjl, 528, L97,
\doi{10.1086/312684} 
%-L100 (HKKM00),  A theoretical light curve model for the 1999 outburst
% of U Scorpii

%\bibitem[Hachisu et al. (2006)]{hac06b}
%Hachisu, I., Kato, M., Kiyota, S., et al. 2006, \apjl, 651, L141
%%% (astro-ph/0607650)
% The Hydrogen Burning Turn-off of RS Ophiuchi 2006

\bibitem[I. Hachisu et al. (2007)]{hac07kl}
Hachisu, I., Kato, M., \& Luna, G. J. M. 2007, \apjl, 659, L153,
\doi{10.1086/516838}
%-L156. Supersoft X-Ray Light Curve of RS Ophiuchi (2006)

%\bibitem[Hachisu et al. (2024)]{hac24km}
%Hachisu, I., Kato, M., \& Matsumoto, K. 2024, \apj, 965, 49,
%\doi{10.3847/1538-4357/ad2a45}
%-19pp.  A multiwavelength light-curve model of the classical nova V339 Del:
% the coexistence of dust dip and supersoft X-rays

%\bibitem[Hachisu et al. (1996)Hachisu, Kato, \& Nomoto]{hkn96}
%Hachisu, I., Kato, M., \& Nomoto, K. 1996, \apj, 470, L97 %(HKN96)
%-L100,  A New Model for Progenitor Systems of Type IA Supernovae

%\bibitem[Hachisu et al. (1999a)]{hknu99}
%Hachisu, I., Kato, M., Nomoto, K., \& Umeda, H.  1999a, \apjl, 519, 314
% He-rich accretion : U Sco type

%\bibitem[Hachisu et al. (1999b)]{hkn99}
%Hachisu, I., Kato, M., \& Nomoto, K. 1999b, \apj, 522, 487

%\bibitem[Hachisu et al. (2010)]{hkn10}
%Hachisu, I., Kato, M., \& Nomoto, K. 2010, \apj, 724, L212
%-L216.  Supersoft X-ray Phase of Single Degenerate Type Ia Supernova
%  Progenitors in Early-type Galaxies

%\bibitem[Hachisu et al. (2012a)]{hksn12}
%Hachisu, I., Kato, M., Saio, H., \&\ Nomoto, K. 2012a, \apj, 744, 69
%, 15 pp. A Single Degenerate Progenitor Model for Type Ia Supernovae
%  Highly Exceeding the Chandrasekhar Mass Limit

%\bibitem[Hachisu et al. (2012b)]{hkn12}
%Hachisu, I., Kato, M., \& Nomoto, K. 2012b, \apj, 56, L4
% 5 pp.  Final Fates of Rotating White Dwarfs and Their Companions
%  in the Single Degenerate Model of Type Ia Supernovae

%\bibitem[Hachisu et al. (2003)]{hac03ks}
%Hachisu, I., Kato, M., \& Schaefer, B. E. 2003, \apj, 584, 1008
%-1015, Revised analysis of the supersoft X-ray phase, helium enrichment,
%and turn-off time in the 2000 outburst of the recurrent nova CI Aquilae
%

\bibitem[I. Hachisu et al. (2025)]{hac25kw}
Hachisu, I., Kato, M., \& Walter, F. 2025, \apj,  980, 142,
\doi{10.3847/1538-4357/adae08}
%in press, arXiv:2412.00250
% A multiwavelength light curve analysis of the classical nova KT Eri:
% Optical contribution from a large irradiated accretion disk

%\bibitem[Hachisu et al. (2016)]{hac16sk}
%Hachisu, I., Saio, H., \& Kato, M. 2016, \apj, 824, 22, 
%\doi{10.3847/0004-637X/824/1/22}
%%% (arXiv:1604.02965)
%  Shortest Recurrence Periods of Forced Novae

\bibitem[I. Hachisu et al. (2020)]{hac20skhs}
Hachisu, I., Saio, H., Kato, M., Henze, M., \& Shafter, A. W. 2020,
\apj, 902, 91, \doi{10.3847/1538-4357/abb5fa}
%%% (arXiv:2009.02937)
% A Theory for the Maximum Magnitude versus Rate of Decline
% (MMRD) Relation of Classical Novae

%\bibitem[Han \& Podsiadlowski (2004)]{han04}
%Han, Z., \& Podsiadlowski, Ph. 2004, \mnras, 350, 1301
%-1309, The single-degenerate channel for the progenitors
% of Type Ia supernovae

%\bibitem[Hayashi et al. (1962)]{hhs62}
%Hayashi, C., Hoshi, R., \& Sugimoto, D. 1962, Prog. Theor. Phys. Suppl., 22. 1
% HHS

%\bibitem[Henze et al. (2014)]{hen14}
%Henze, M., Ness, J.-U., Darnley, M., et al. 2014, \aap, 563, L8
%(arXiv1401.2904)
% Henze, M.; Ness, J.-U.; Darnley, M. J.; Bode, M. F.; Williams, S. C.;
% Shafter, A. W.; Kato, M.; Hachisu, I.
% A remarkable recurrent nova in M31 - The X-ray observations

%\bibitem[Henze et al. (2015a)]{hen15a}
%Henze, M., Ness, J.-U., Darnley, M. J., et al. 2015a, \aap, 580, A46 
%12 pp.  A remarkable recurrent nova in M 31: 
%  The predicted 2014 outburst in X-rays with Swift


%\bibitem[Henze et al. (2015b)]{hen15b}
%Henze, M., Darnley, M. J., Kabashima, F.,  et al.
%%%Nishiyama, K., Itagaki, K., \& Gao, X. 
%2015b, \aap, 582, L8 
%4 pp. A remarkable recurrent nova in M 31: The 2010 eruption recovered
% and evidence of a six-month period

\bibitem[M. Henze et al. (2018)]{hen18}
Henze, M., Darnley, M. J., Williams, S. C., et al. 2018, \apj, 857, 68,
\doi{10.3847/1538-4357/aab6a6}
%, 29 pp. (2018).  Breaking the Habit: The Peculiar 2016 Eruption of 
% the Unique Recurrent Nova M31N 2008-12a

%\bibitem[Hillman et al. (2014)]{hil14}
%Hillman, Y., Prialnik, D., Kovetz, A., Shara, M. M., \& Neill, J. D. 
%2014, \mnras, 437, 1962
% Nova multiwavelength light curves: predicting UV precursor flashes
% and pre-maximum halts

%\bibitem[Hillman et al. (2015)]{hil15}
%Hillman, Y., Prialnik, D., Kovetz, A., \& Shara, M. M. 2015, \mnras, 446, 1924
% Observational signatures of SNIa progenitors, as predicted by models

%\bibitem[Hillman et al. (2016)]{hil16}
%Hillman, Y., Prialnik, D., Kovetz, A., \& Shara, M. M. 2016, \apj, 819, 168
% He flash model ;wrong calculation 

\bibitem[Y. Hillman et al. (2019)]{hil19}
Hillman, Y., Orio, M., Prialnik, D., et al. 2019, \apjl, 879, L5,
\doi{10.3847/2041-8213/ab2887}
%, 5 pp. The Supersoft X-Ray Transient ASASSN-16oh 
% as a Thermonuclear Runaway without Mass Ejection

\bibitem[R. Hounsell et al. (2010)]{hou10bh}
Hounsell, R., Bode, M. F., Hick, P. P., et al. 2010, \apj, 724, 480,
\doi{10.1088/0004-637X/724/1/480}
%-486.  Exquisite Nova Light Curves from the Solar Mass Ejection Imager (SMEI)

%\bibitem[Iben (1982)]{ibe82}
%Iben, I., Jr. 1982,  \apj, 259, 244
% hot accreting WDs in the quasi-static approximation

%\bibitem[Idan et al. (2013)]{ida13}
%Idan, I., Shaviv, N. J., \& Shaviv, G. 2013, \mnras, 433, 2884
% The fate of a WD accreting H-rich material at high accretion rates

%\bibitem[Iglesias \& Rogers (1996)]{igl96}
%Iglesias, C. A., \& Rogers, F. J. 1996, \apj, 464, 943
%-953. Updated Opal Opacities

%\bibitem[Imamura \& Tanabe (2012)]{ima12t}
%Imamura, K., \& Tanabe, K. 2012, \pasj, 64, 120, \doi{10.1093/pasj/64.6.120}
%%Imamura, Kazuyoshi; Tanabe, K. 2012, \pasj, 64, 120
% (8 pp.) Observational Report on the Classical Nova KT Eridani

%\bibitem[Imara \& Blitz (2007)]{imara07}
%Imara, N., \& Blitz, L. 2007, \apj, 662, 969
%-979.  Extinction in the Large Magellanic Cloud

\bibitem[S. W. Jha et al. (2016)]{jha16}
Jha, S. W., Colmenero, E. R., Stanek, K. Z., et al. 2016,
The Astronomer's Telegram, 9859, 1  
% ASASSN-16oh: An Unusual Transient in the Vicinity of the SMC

%\bibitem[Kahabka \& van den Heuvel (1997)]{kah97}
%Kahabka, P., \& van den Heuvel, E. P. J. 1997, \araa, 35, 69
%%-100.  Luminous Supersoft X-Ray Sources

% \bibitem[Kato (1983)]{kat83}
% Kato, M. 1983, \pasj, 35, 507
% Optically  Thick Winds and Nova Outbursts

%\bibitem[Kato (1985)]{kat85} 
%Kato, M. 1985, \pasj, 37, 19
% Occurrence of Optically  Thick Winds

%\bibitem[Kato (1997)]{kat97} 
%Kato, M. 1997, \apjs, 113, 121
% optically thick winds from degenerate dwarfs I.CN of population I and II 

\bibitem[M. Kato (1999)]{kat99} 
Kato, M. 1999, \pasj, 51, 525, \doi{10.1093/pasj/51.4.525}
% Optically thick wind solutions for an extremely rapid light curve of RNe
%Publications of the Astronomical Society of Japan, v.51, p.525-535.
% Pub Date: August 1999 DOI: 10.1093/pasj/51.4.525

%\bibitem[Kato (2013)]{kat13}
%Kato, M. 2013, ``Binary Paths to Type Ia supernovae Explosions'', 
%IAU Symp. 281,  (Cambridge: Cambridge University Press) 172 
% Novae and Accreting White Dwarfs as Progenitors of Type Ia Supernovae

% \bibitem[Kato \& Hachisu (1991)]{kat91}
% Kato, M., \& Hachisu, I., 1991, \apj, 373, 620
% Is Drag Luminosity Effective in Recurrent Novae?

%\bibitem[Kato \& Hachisu (1994)]{kat94h}
%Kato, M., \& Hachisu, I., 1994, \apj, 437, 802, \doi{10.1086/175041}
%-826  Optically thick winds in nova outbursts.

%\bibitem[Kato \& Hachisu (2004)]{kat04}
%Kato, M., \& Hachisu, I., 2004, \apj, 613, L129
% Mass accumulation efficiency in He shell flashes for various WD masses 

%\bibitem[Kato \& Hachisu (2009)]{kat09}
%Kato, M., \& Hachisu, I., 2009, \apj, 699, 1293
% Multiplicity of optically thick winds 

%\bibitem[Kato \& Hachisu (2011b)]{kat11b}
%Kato, M., \& Hachisu, I., 2011b, \apj, 743, 157
% Effects of a companion star on slow nova outbursts-transition
% from static to wind evolution

%\bibitem[Kato \& Hachisu (2012)]{kat12review}
%Kato, M., \& Hachisu, I. 2012, BASI, 40, 393
%%%Bulletin of the Astronomical Society of India, 40, 393
% -417 Recurrent novae as a progenitor of Type Ia supernovae

%\bibitem[Kato et al. (2011)]{kat11}
%Kato, M., Hachisu, I., Cassatella, A., \& Gonz\'alez-Riestra, R.
%2011, \apj, 727, 72
% PU Vul model

%\bibitem[Kato et al. (2009)]{kat09}
%Kato, M., Hachisu, I., \& Cassatella, A., 2009, \apj, 704, 1676
%-1688.  A Universal Decline Law of Classical Novae. IV.  V838 Her (1991):
% A Very Massive White Dwarf

%\bibitem[Kato et al. (2013)]{kat13hh}
%Kato, M., Hachisu, I., \& Henze, M.  2013, \apj, 779, 19
% novae in globular clusters

%\bibitem[Kato et al. (2008)]{kat08}
%Kato, M., Hachisu, I., Kiyota, S. \& Saio, H. 2008, \apj, 684, 1366
% V445 Pup

%\bibitem[Kato et al. (2017c)]{kat17palermo}
%Kato, M.,  Hachisu, I., \& Saio, H. 2017c, 
%in Proceedings of the Palermo Workshop
%2017 on ``The Golden Age of Cataclysmic Variables and Related
%Objects - IV'', ed. F. Giovannelli et al. (Trieste: SISSA PoS), 315, 56
% Palermo conference: compare numerical codes

%\bibitem[Kato et al. (2012)]{kat12mh}
%Kato, M., Miko\l ajewska, J., \& Hachisu, I., 2012, \apj, 750, 5
% PU Vul

%\bibitem[Kato et al. (2014)]{kat14shn}
%Kato, M., Saio, H., Hachisu, I., \& Nomoto, K. 2014, \apj, 793, 136,
%\doi{10.1088/0004-637X/793/2/136}
% shortest recurrence periods of novae

%\bibitem[Kato et al. (2015)]{kat15sh}
%Kato, M., Saio, H., \& Hachisu, I. 2015, \apj, 808, 52, 
%4 pp.  Multi-wavelength Light Curve Model
% of the One-year Recurrence Period Nova M31N 2008-12A

%\bibitem[Kato et al. (2017a)]{kat17sha}
%Kato, M., Saio, H., \& Hachisu, I. 2017a, ApJ, 838, 153
% A Self-consistent Model for a Full Cycle of RNe
% 1.2 Mo and 1.38 Mo  :HR diagram

%\bibitem[Kato et al. (2017b)]{kat17shb}
%Kato, M., Saio, H., \& Hachisu, I. 2017b, ApJ, 844, 143
% A Millennium-long Evolution of 1 yr Recurrence Period Nova
% -- Search for Any Indication of the Forthcoming He Flash

\bibitem[M. Kato et al. (2020)]{kat20sh}
Kato, M., Saio, H., \& Hachisu, I. 2020, \apj, 892, 15,
\doi{10.3847/1538-4357/ab7996}
% ASASSN-16oh: A Nova Outburst with No Mass Ejection¡½A New Type of
% Supersoft X-Ray Source in Old Populations
% Kato, Mariko; Saio, Hideyuki ; Hachisu, Izumi
% The Astrophysical Journal, Volume 892, Issue 1, id.15, 16 pp. (2020)
% Pub Date: March 2020 DOI: 10.3847/1538-4357/ab7996 

\bibitem[M. Kato et al. (2021)]{kat21sh}
Kato, M., Saio, H, \& Hachisu, I. 2021, \pasj, 73, 1137,
\doi{10.1093/pasj/psab0764}
% model V2491 Cyg, H ignition for massive WD

\bibitem[M. Kato et al. (2022a)]{kat22sha}
Kato, M., Saio, H., \& Hachisu, I. 2022a, \pasj, 74, 1005,
\doi{10.1093/pasj/psac051}
%in press (arXiv:2206.03136)
% Physics of nova outbursts - A theoretical model of
% classical nova outbursts with self-consistent wind mass loss
% Publications of the Astronomical Society of Japan, Volume 74,
% Issue 5, pp.1005-1021   Pub Date: October 2022 DOI: 10.1093/pasj/psac051

\bibitem[M. Kato et al. (2022b)]{kat22shb}
Kato, M., Saio, H., \& Hachisu, I. 2022b, \apjl, 935, L15,
\doi{10.3847/2041-8213/ac85c1}
%%%(arXiv:2206.03136)
% A Light-curve Analysis of the X-Ray Flash First Observed in Classical Novae
% The Astrophysical Journal Letters
% 2022-08-01 | Journal article
% DOI: 10.3847/2041-8213/ac85c1
% CONTRIBUTORS: Mariko Kato; Hideyuki Saio; Izumi Hachisu

\bibitem[M. Kimura et al. (2018)]{kim18}
Kimura, M., Kato, T., Maehara, H., et al.
%%%Ishioka, R., Berto, M. et al. 
2018, \pasj, 709, 78, \doi{10.1093/pasj/psy073} 
%Kimura, Mariko; Kato, Taichi; Maehara, Hiroyuki; Ishioka, Ryoko; Monard, Berto; Nakajima, Kazuhiro; Stone, Geoff; Pavlenko, Elena P.; Antonyuk, Oksana I.; Pit, Nikolai V.; and 28 coauthors
%On the nature of long-period dwarf novae with rare and low-amplitude outbursts

%\bibitem[Kovetz (1998)]{kov98}
%Kovetz, A. 1998, \apj, 495, 401
%-406. The Simulation of Nova Evolution with Optically Thick Winds

%\bibitem[Langer et al. (2000)]{lan00}
%Langer, N., Deutschmann, A., Wellstein, S., \& H\"oflich, P. 2000, \aap, 362, 1046
%-1064. The evolution of main sequence star + white dwarf binary systems
% towards Type Ia supernovae

\bibitem[J.-P. Lasota (2001)]{las01}
Lasota, J.-P. 2001, New Astronomy Reviews, 45, 449,
\doi{10.1016/S1387-6473(01)00112-9}
%The disc instability model of dwarf novae and low-mass X-ray binary transients
%Lasota, Jean-Pierre
%New Astronomy Reviews, Volume 45, Issue 7, p. 449-508.
% Pub Date: June 2001 DOI: 10.1016/S1387-6473(01)00112-9 

%\bibitem[Li \& van den Heuvel (1997)]{li97}
%Li, X.-D., \& van den Heuvel, E. P. J. 1997, \aap, 322, L9
%-L12.  Evolution of white dwarf binaries: supersoft X-ray sources
% and progenitors of type IA supernovae.

%\bibitem[Liu et al. (2019)]{liu19}
%Liu, J., Zhang, H., Howard, A. W. 2019, \nat, 575, 618
% A wide star -black-hole binary system from radial-velocity 
% measurements

%\bibitem[Liszt (2014)]{lis14}
%Liszt, H. S. 2014, \apj, 780, 10 %%% in press (arXiv:1310.6616)
%-522. N(H I)/E(B-V)

%\bibitem[Lockley et al. (1997)Lockley, Eyres, \& Wood]{loc97}
%Lockley, J. J., Eyres, S. P. S., \& Wood, Janet H. 1997, \mnras, 287, L14
%-L16, A radio detection of V Sagittae

%\bibitem[Lockwood \& Millis (1976)]{loc76m}
%Lockwood, G. W., \& Millis, R. L. 1976, \pasp, 88, 235, \doi{10.1086/129935}
%-237. UVBY light curves of Nova CYG 1975.

%\bibitem[Ma et al. (2013)]{ma13}
%Ma, X., Chen, X., Chen, H-L., Denissenkov, P. A., \& Han, Z.  2013, 
%\apjl, 778, L32
% A super-Eddington wind scenario for the progenitors of type Ia SNe
 
% \bibitem[Maccarone et al. (2016)]{mac16} xxxx ATel 9866

\bibitem[T. J. Maccarone et al. (2019)]{mac19}
Maccarone, T. J., Nelson, T. J., Brown, P. J., et al. 2019, 
Nature Astronomy, 3, 173, \doi{10.1038/s41550-018-0639-1}
%(arXiv:1907.02130)
% -177.  Unconventional origin of supersoft X-ray emission 
% from a white dwarf binary 

%\bibitem[MacDonald \& Vennes (1991)]{mac91}
%MacDonald. J. \& Vennes, S. 1991, \apj, 373, L51
%Thermal X-ray emission from CNe in optical decline 


%\bibitem[Maoz et al. (2014)]{mao14}
%Maoz, D., Mannucci, F., \& Nelemans, G. 2014, \araa, 52, 107
% (arXiv:1312.0628)
%%Annual Reviews of Astronomy and Astrophysics, 2014
% Observational clues to the progenitors of Type-Ia supernovae

\bibitem[P. Mr\'oz et al. (2016)]{mro16}
Mr\'oz, P., Udalski, A., Wyrzykowski, L., Kozlowski, S., \& Poleski, R.
2016, The Astronomer's Telegram, 9867, 1
%  OGLE-IV Observations of ASASSN-16oh

\bibitem[P. Mr\'oz et al. (2024)]{mro24ks}
Mr\'oz, P., Kr\'ol, K., Szegedi, K., et al. 2024, \apjl, 977, L37,
\doi{10.3847/2041-8213/ad969b}
%(arXiv2409.17338) 
%  Millinovae: A New Class of Transient Supersoft X-ray Sources
%  without a Classical Nova Eruption
% 977, Issue 2, id.L37, 13 pp.
 
%\bibitem[Munari et al. (2013a)]{mun13dcvf}
%Munari, U., Dallaporta, S., Castellani, F., et al. 2013a, \mnras, 435, 771,
%\doi{10.1093/mnras/stt1340}
%-781. Photometric evolution, orbital modulation and progenitor
%  of Nova Mon 2012 (V959 Mon)

%\bibitem[Munari et al. (2015)]{mun15mm}
%Munari, U., Maitan, A., Moretti, S., Tomaselli, S. 2015, NewA, 40, 28,
%\doi{10.1016/j.newast.2015.03.008}
% 500 days of Stromgren b, y and narrow-band [OIII], H_alpha
% photometric evolution of gamma-ray Nova Del 2013 (=V339 Del)

%\bibitem[Muraoka et al. (2024)]{mura24ki}
%Muraoka, K., Kojiguchi, N., Ito, J., et al. 2024, \pasj, 76, 293,
%\doi{10.1093/pasj/psae010}
%Optical and soft X-ray light-curve analysis during the 2022 eruption
% of U Scorpii: Structural changes in the accretion disk

%\bibitem[Nariai et al. (1980)]{nar80}
%Nariai, K., Nomoto, K., \& Sugimoto, D. 1980, \pasj, 32, 473
% -494  Nova explosion of Mass-accreting WDs. 1.3 and 0.4 Mo

%\bibitem[Nelson et al. (2011)]{nel11}
%Nelson, T., Mukai, K., Orio, M., Luna, G.J.M., Sokoloski, J., L.
%2011, \apj, 737, 7
% X-ray observation in quiescent phase of RS Oph

%\bibitem[Neo et al. (1977)]{neo77}
%Neo, S., Miyaji, S., Nomoto, K., \& Sugimoto, D. 1977, \pasj, 29, 249

%\bibitem[Nomoto (1982)]{nom82}
%Nomoto, K. 1982, \apj, 253, 798
%-810.  Accreting white dwarf models for type I supernovae.
%  I - Presupernova evolution and triggering mechanisms

%\bibitem[Nomoto et al. (2007)]{nom07}
%Nomoto, K., Saio, H., Kato, M., \& Hachisu, I. 2007, \apj, 663, 1269
% Thermal stability of WDs accreting H-rich matter and progenitors of SN Ia

%\bibitem[Ootsuki et al. (2009)]{oot09ow}
%Ootsuki, I., Ohshima, O., Watanabe, O., et al. 2009, \iaucirc, 9098, 2
% KT Eridani=Nova Eridani 2009

%\bibitem[Orio et al. (2013)]{ori13}
%Orio, M., Behar, E., Gallagher, J., et al. 2013, \mnras, 429, 1342
%-1353. Thomson scattering and collisional ionization in the X-ray
% grating spectra of the recurrent nova U Scorpii

\bibitem[Y. Osaki (1996)]{osa96}
Osaki, Y. 1996, \pasp, 108, 39, \doi{10.1086/133689}
% dwarf nova outburst:review

%\bibitem[Paczy\'nski \& \.Zytkow (1978)]{pac78}
% Paczy\'nski, B., \& \.Zytkow, A. N. 1978, \apj, 222, 604
%-611. Hydrogen shell flashes in a white dwarf with mass accretion

\bibitem[B. Paczy\'nski \& A. Schwarzenberg-Czerny (1980)]{pac80s}
Paczy\'nski, B., \& Schwarzenberg-Czerny, A.
1980, Acta Astronomica, 30, 127
%Disk accretion in U Gem.
%Acta Astron., Vol. 30, p. 127-141 (1980)
% Pub Date: 1980


%\bibitem[Page et al. (2010)]{pag10}
%Page, K.,L., Osborne, J. P., Evans, P.A. et al \mnras, 2010, 401, 121
% Swift observation of the X-ray and UV evolution of V2491 Cyg

%\bibitem[Pagnotta \& Schaefer (2014)]{pag14}
%Pagnotta, A., \& Schaefer, B. E. 2014, \apj, 788, 164
% Identifying and Quantifying Recurrent Novae Masquerading as Classical Novae
% after that Brad claims all the accreting WDs never be SNe Ia.

%\bibitem[Pagnotta et al. (2015)]{pagnotta15}
%Pagnotta, A., Schaefer, B. E., Clem, J. L., et al. 2015, \apj, 811, 32,
%\doi{10.1088/0004-637X/811/1/32}
% 42 pp.  The 2010 Eruption of the Recurrent Nova U Scorpii:
%   The Multi-wavelength Light Curve

%\bibitem[Pakull et al. (1993)]{pak93}
%Pakull, M. W., Moch, C., Bianchi, L., et al.
%%Thomas, H.-C., Guibert, J., Beaulieu, J. P., Grison, P., \& Schaeidt, S. 
%1993, \aap, 278, L39
%-L42, Optical/UV counterpart of the supersoft transient X-ray source
% RX J0513.9-6951 in the Large Magellanic Cloud

%\bibitem[Parikh et al. (2019)]{par19}
%Parikh, A. S., Hern\'andez Santisteban, J. V., Wijnands, R., Page, D.
%2019, Revista Mexicana de Astronomia y Astrofisca, 55, 55
%%%2019RMxAA..55...55P
%Multiwavelength Observations of MASTER OT 075353.88+174907.6: 
%A Likely Superoutburst of a Long Period Dwarf Nova System

%\bibitem[Patterson et al. (1998)]{pat98}
%Patterson, J. et al. 1998, \pasp, 110, 380
%-395, Two Galactic Supersoft X-Ray Binaries: V Sagittae and T Pyxidis

\bibitem[J. Patterson (2011)]{pat11}
Patterson, J. 2011, \mnras, 411, 2695, \doi{10.1111/j.1365-2966.2010.17881.x}
%Distances and absolute magnitudes of dwarf novae: murmurs of period bounce
%Patterson, Joseph
%Monthly Notices of the Royal Astronomical Society, Volume 411,
% Issue 4, pp. 2695-2716.
% Pub Date: March 2011 DOI: 10.1111/j.1365-2966.2010.17881.x

%\bibitem[Patterson et al. (2014)]{pat14}
%Patterson, J., Oksanen, A., Monard, B., et al. 2014, Stella Novae:
%Past and Future Decades, ed. P. A. Woudt \& V. A. R. M. Ribeiro
%(San Francisco, CA, ASP), ASP Conf. Ser. 490, 35
% The Death Spiral of T Pyxidis


%\bibitem[Pietrzy\'nski et al. (2013)]{pie13}
%Pietrzy\'nski, G., Graczyk, D., Gieren, W., et al. 2011, \nat, 495, 76
%-79 (2013).  An eclipsing-binary distance to
% the Large Magellanic Cloud accurate to two per cent

%\bibitem[Piro \& Bildsten (2004)]{pir04}
%Piro, A.L., \& Bildsten, L. 2004, \apj, 610,979
% spreading layer 

%\bibitem[Popham \& Di Stefano (1996)]{pop96}
%Popham, R., \& Di Stefano, R. 1996, Lecture Notes in Physics, 472, p.66,
%ed. Greiner, J. (Springer:Berlin)
% accretion disks in supersoft X-ray sources

%\bibitem[Popham \& Narayan (1995)]{pop95}
%Popham, R., \& Narayan, R. 1995, \apj, 442, 337
% Accretion disk boundary layers in CV.I. optically thick boundary layer
 
%\bibitem[Prialnik (1986)]{pri86}
%Prialnik, D. 1986, \apj, 310, 222
% The evolution of a classical nova model through a complete cycle

%\bibitem[Prialnik \& Kovetz (1995)]{pri95}
%Prialnik, D., \& Kovetz, A. 1995, \apj, 445, 789
% grid of nova cycles

\bibitem[E. Ragan et al. (2009)]{rag09bs}
Ragan, E., Brozek, T., Suchomska, K., et al. 2009, ATel, 2327, 1
% Optical observations of KT Eri = Nova Eridani 2009

%\bibitem[Raj et al. (2012)]{raj12}
%Raj, A., Ashok, N. M., Banerjee, D. P. K., et al.
%%%%Munari, U., Valisa, P., \& Dallaporta, S.
%2012, \mnras, 425, 2576
%-2588. V496 Scuti: an Fe II nova with dust shell accompanied by CO emission

\bibitem[A. Rajoelimanana et al. (2017)]{raj17cb}
%Multi-wavelength observations of the unusual soft X-ray transient ASASSN-16oh
Rajoelimanana, A., Charles, P., Buckley, D., \& Meintjes, P. 2017,
% Publication:  
in 5th Annual Conference on High Energy Astrophysics in Southern Africa,
%%%4-6 October 2017. University of the Witwatersrand (Wits), South Africa.
eds. P. Meintejis et al., (SISSA: Trieste, Italy), 319, 3,
%%%Online at https://pos.sissa.it/cgi-bin/reader/conf.cgi?confid=319, id.3,
\doi{10.22323/1.319.0003}
%%% Pub Date: October 2017 DOI: 10.22323/1.319.0003 

%\bibitem[Rieke \& Lebofsky (1985)]{rie85}
%Rieke, G. H., \& Lebofsky, M. J. 1985, \apj, 288, 618, \doi{10.1086/162827}
%-621, The interstellar extinction law from 1 to 13 microns

%\bibitem[Reinsch et al. (1996)]{rei96}
%Reinsch, K., van Teeseling, A., Beuermann, K., \& Abbott, T. M. C. 1996,
%\aap, 309, L11
%-L14, Optical low states of the supersoft X-ray source RX J0513.9-6951

%\bibitem[Reinsch et al. (2000)]{rei00}
%Reinsch, K., van Teeseling, A., King, A. R., \& Beuermann, K. 2000,
%\aap, 354, L37
%-L40, A limit-cycle model for the binary supersoft X-ray source
%RX J0513.9-6951

\bibitem[A. C. Rodriguez et al. (2024)]{rod24es}
Rodriguez, A. C., El-Badry, K., Suleimanov, V., et al. 2024,
%Cataclysmic Variables and AM CVn Binaries in SRG/eROSITA + Gaia:
% Volume Limited Samples, X-ray Luminosity Functions, and Space Densities
arXiv:2408.16053
% Pub Date: August 2024 DOI: 10.48550/arXiv.2408.16053

%\bibitem[Sahman et al. (2013)]{sah13}
%Sahman, D. I., Dhillon, V. S., Marsh, T. R., et al. 2013, \mnras, 433, 1588
%-1598.  CI Aql: a Type Ia supernova progenitor?

\bibitem[I. V. Salazar et al. (2017)]{salazar17}
Salazar, I. V., LeBleu, A., Schaefer, B. E., Landolt, A. U., \& Dvorak, S.
2017, \mnras, 469, 4116, \doi{10.1093/mnras/stx1161}
%-4132.  Accurate pre- and post-eruption orbital periods 
% for the dwarf/classical nova V1017 Sgr

%\bibitem[Schaefer (2001)]{sch01c}
%Schaefer, B. E. 2001, \iaucirc\  7750
% CI AQUILAE  --- the 1941 possible outburst

%\bibitem[Schaefer et al. (2013)]{sch13}
%Schaefer, B. E., Landolt, A. U., Linnolt, M., et al. 2013, \apj, 773, 55
% 23 pp. The 2011 Eruption of the Recurrent Nova T Pyxidis:
% The Discovery, the Pre-eruption Rise, the Pre-eruption Orbital Period,
% and the Reason for the Long Delay

%\bibitem[Schaefer (2018)]{schaefer18}
%Schaefer, B. E. 2018, \mnras, 481, 3033
%-3051.  The distances to Novae as seen by Gaia

%\bibitem[Schaefer (2022a)]{schaefer22a}
%Schaefer, B.E., 2022a, \mnras, 517, 3640, \doi{10.1093/mnras/stac2089}
% Comprehensive listing of 156 reliable orbital periods of nova
% including 49 new periods
%  Res Notes AAS, 5, 150;  (arXiv:2207.02932)
%Discovery of 13 New orbital periods for CNe

%\bibitem[Schaefer (2022b)]{schaefer22b}
%%Comprehensive catalogue of the overall best distances and properties
%% of 402 galactic novae
%Schaefer, B. E. 2022b, \mnras, 517, 6150, \doi{10.1093/mnras/stac2900}
%Monthly Notices of the Royal Astronomical Society, Volume 517, Issue 4,
% pp.6150-6169
% Pub Date: December 2022 DOI: 10.1093/mnras/stac2900

%\bibitem[Schaefer \& Ringwald (1995)]{sch95r}
%Schaefer, B. E., \& Ringwald, F. A. 1995, \apjl, 447, L45
%-L48.  Improved Orbital Period for the Recurrent Nova U Scorpii

\bibitem[B. E. Schaefer et al. (2022)]{schaefer22wh}
Schaefer, B. E., Walter, F. M., Hounsell, R., Hillman, Y. 2022, \mnras,
517, 3864, \doi{10.1093/mnras/stac2923}
%  The nova KT Eri is a recurrent nova with a recurrence time-scale of
%  40-50 yr.

%\bibitem[Schaeidt et al. (1993)Schaeidt, Hasinger, \& Truemper]{sch93}
%Schaeidt, S., Hasinger, G., \& Truemper, J. 1993, \aap, 270, L9
%-L12, Discovery of a variable supersoft X-ray source in the Large
% Magellanic Cloud during the ROSAT All-Sky Survey

%\bibitem[Schandl et al. (1997)]{sch97}
%Schandl, S., Meyer-Hofmeister, E., \& Meyer, F. 1997, \aap, 318, 73
%Schandl, S., Meyer-Hofmeister, E. \& Meyer, F.
%Visual light from the eclipsing supersoft X-ray source CAL 87.
%{\it Astron. Astrophys.}  318, 73---80 (1997).

%\bibitem[Schwarzschild (1958)]{sch58}
%Schwarzschild, M. 1958, 
%Structure and Evolution of the Stars, (New York: Dover)

\bibitem[A. D. Schwope et al. (2024)]{schwope24kk}
Schwope, A. D., Knauff, K., Kurpas, J., et al. 2024, \aap, 690, A243,
\doi{10.1051/0004-6361/202450537}
% A first systematic characterization of cataclysmic variables
% in SRG/eROSITA surveys
%Astronomy & Astrophysics, Volume 690, id.A243, 20 pp.
% Pub Date: October 2024 DOI: 10.1051/0004-6361/202450537 

%\bibitem[Schwarz et al. (2011)]{schw11}
%Schwarz, G. J., Ness, J.-U., Osborne, J. P., et al. 2011, \apjs, 197, 31
% 26 pp.  Swift X-Ray Observations of Classical Novae.
%  II. The Super Soft Source Sample

%\bibitem[Shara et al. (1986)]{sha86}
%Shara, M. M., Livio, M., Moffat, A. F. J., \& Orio, M. 1986, \apj, 311, 163
%-171. Do novae hibernate during most of the millennia between eruptions?
% Links between dwarf and classical novae, and implications
% for the space densities and evolution of cataclysmic binaries

%\bibitem[Shen \& Bildsten (2007)]{she07}
%Shen, K. J., \& Bildsten, L. 2007, \apj, 660, 1444
%-1450.  Thermally Stable Nuclear Burning on Accreting White Dwarfs

%\bibitem[Shen \& Bildsten (2008)]{she08}
%Shen, K. J., \& Bildsten, L. 2008, \apj, 668, 1530 (Erratum)
%  Thermally Stable Nuclear Burning on Accreting White Dwarfs

%\bibitem[Shen et al. (2009)]{she09}
% Shen, K. J. Idan, I., \& Bildsten, L.  2009, \apj, 705, 693
% Helium Core White Dwarfs in Cataclysmic Variables

%\bibitem[Sienkiewics (1975)]{sie75}
%Sienkiewicz, R. 1975, \aap, 45, 411

%\bibitem[Sienkiewics (1980)]{sie80}
%Sienkiewicz, R. 1980, \aap, 85, 295

%\bibitem[\v{S}imon (1996a)]{sim96a}
%\v{S}imon, V. 1996a, \aaps, 118, 421
%-428, The peculiar interacting binary V Sagittae:
% Brightness variations in 1932-1994

%\bibitem[\v{S}imon (1996b)]{sim96b}
%\v{S}imon, V. 1996b, \aap, 309, 775
%-776, Possible change of the orbital period
% of the nova-like binary V Sagittae

%\bibitem[\v{S}imon et al. (2002)]{sim02}
%\v{S}imon, V., Hric, L., Petr\'{\i}c, K., Shugarov, S.,
% Niarchos, P., \& Marsakova, V. I. 2002, \aap, 393, 921
%-925, The orbital modulation of the X-ray binary V Sagittae
% in the high and low states

%\bibitem[\v{S}imon \& Mattei (1999)]{sim99}
%\v{S}imon, V., \& Mattei, J. A. 1999, \aaps, 139, 75
%-88, The peculiar binary V Sagittae: Properties of its long-term
% light changes

%\bibitem[\v{S}imon et al. (2001)]{sim01}
%\v{S}imon, V., Shugarov, S., Marsakova, V. I. 2001, \aap, 366, 100
%-105, Long-term color variations of the peculiar X-ray binary V Sagittae

%\bibitem[Sion et al. (1979)]{sio79}
%Sion, E. M., Acierno, M. J., \& Tomczyk, S. 1979, \apj, 230, 832
% Hydrogen shell flashes in massive accreting white dwarfs

% \bibitem[Starrfield et al. (1998)]{sta98}
% Starrfield, S., Truran, J. W., Wiescher, M. C., \& Sparks, W. M.
% 1998, \mnras, 296, 502
% Evolutionary sequences for Nova V1974 Cygni using new nuclear reaction rates and opacities

%\bibitem[Smak (1995)]{sma95}
%Smak, J. I. 1995, Acta Astr., 45, 361
%-364, On the Orbital Period of V Sge

%\bibitem[Smak et al. (2001)Smak, Belczynski, \& Zola]{sma01}
%Smak, J. I., Belczynski, K., \& Zola, S. 2001, Acta Astr., 51, 117
%-150, V Sge: a Hot, Peculiar Binary System

%\bibitem[Southwell et al. (1996)]{sou96}
%Southwell, K. A., Livio, M., Charles, P. A., O'Donoghue, D.,
%\& Sutherland, W. J. 1996, \apj, 470, 1065
%-1074, The Nature of the Supersoft X-Ray Source RX J0513-69

%\bibitem[Starrfield et al. (2004)]{sta04}
%Starrfield, S., Timmes, F. X., Hix, W. R., et al.
%%% Sion, E. M., Sparks, W. M., Dwyer, S. J.
%2004, \apjl, 612, L53
%%-L56.  Surface Hydrogen-burning Modeling of Supersoft X-Ray Binaries:
%%  Are They Type Ia Supernova Progenitors?

%\bibitem[Starrfield et al. (2012)]{sta12bal}
%Starrfield, S., Timmes, F. X., Iliadis, C., et al. 2012, Baltic Astron., 21, 76
%%% Hix, W. R., Arnett, W. D., Meakin, C., Sparks, W. M.
%%-87.  Hydrodynamic Studies of the Evolution of Recurrent, Symbiotic
%%  and Dwarf Novae: the White Dwarf Components are Growing in Mass

%\bibitem[Starrfield et al. (2012b)]{sta12}
%Starrfield, S., Iliadis, C., Timmes, F. X., et al. 2012b, BASI, 40, 419
%%Hix, W. R., Arnett, W. D., Meakin, C., Sparks, W. M.
%%  Theoretical studies of accretion of matter onto white dwarfs
%%  and the single degenerate scenario for supernovae of Type Ia

%\bibitem[Steiner \& Diaz (1998)]{ste98}
%Steiner, J. E., \& Diaz, M. P. 1998, \pasp, 110, 276
%-282, The V Sagittae Stars

%\bibitem[Sugimoto (1970)]{sug70}
%Sugimoto, D. 1970, \apj,  159, 619

%\bibitem[Sugimoto \& Fujimoto (1978)]{sug78}
%Sugimoto, D. \& Fujimoto, M. Y. 1978, \pasj, 30,467
% A General Theory for Thermal Pulses of finite amplitude
% in Nuclear shell burnings

%\bibitem[Tang et al. (2014)]{tan14}
%Tang, S., Bildsten, L., Wolf, W. M., et al. 2014, \apj, 786, 61
%(arXiv1401.2426)
% An Accreting White Dwarf near the Chandrasekhar Limit in the Andromeda Galaxy

%\bibitem[Thoroughgood et al. (2001)]{tho01}
%Thoroughgood, T. D., Dhillon, V. S., Littlefair, S. P., Marsh, T. R.,
% \& Smith, D. A. 2001, \mnras, 327, 1323
%-1333. The mass of the white dwarf in the recurrent nova U Scorpii

\bibitem[A. Udalski et al. (2015)]{uda15ss}
Udalski, A., Szyma\'nski, M.K., Szyma\'nski, G. 2015, Acta Astronomica, 65, 1
% OGLE IV 
%OGLE-IV: Fourth Phase of the Optical Gravitational Lensing Experiment
%Udalski, A., Szyma\'nski, M. K., ; Szyma\'nski, G. Acta Astronomica,
% vol 65, no 1, p. 1-38  Pub Date: March 2015 DOI: 10.48550/arXiv.1504.05966 

%\bibitem[Uthas et al. (2010)]{uth10}
%Uthas, H., Knigge, C., \& Steeghs, D. 2010, \mnras, 409, 237
%-246. The orbital period and system parameters of the recurrent nova T Pyx

%\bibitem[van den Heuvel et al. (1992)]{vdh92}
%van den Heuvel, E. P. J., Bhattacharya, D., Nomoto, K., \& Rappaport,
%S. A.  1992, \aap, 262, 97
%%van den Heuvel, E. P. J., Bhattacharya, D., Nomoto, K., \& Rappaport,
%%S. Accreting white dwarf models for CAL 83, CAL 87 and other
%%ultrasoft X-ray sources in the LMC. {\it Astron. Astrophys.}
%% 262, 97---105 (1992).

%\bibitem[Yoon et al. (2004)]{yoo04}
%Yoon, S.-C., Langer, N., \& van der Sluys, M. 2004, \aap, 425, 207
%-216. On the stability of thermonuclear shell sources in stars 

\bibitem[F. M. Walter et al. (2012)]{wal12bt}
Walter, F. M., Battisti, A., Towers, S. E., Bond, H. E.,
\& Stringfellow, G. S. 2012, \pasp, 124, 1057, \doi{10.1086/668404}
%-1072. The Stony Brook/SMARTS Atlas of (mostly) Southern Novae

%\bibitem[Wolf et al. (2013)]{wol13bb}
%Wolf, W. M., Bildsten, L., Brooks, J., \& Paxton, B. 2013, \apj, 777, 136,
%\doi{10.1088/0004-637X/777/2/136}
%(Erratum: 2014, \apj, 782, 117, \doi{10.1088/0004-637X/782/2/117})
% Hydrogen Burning on Accreting White Dwarfs: Stability, Recurrent Novae,
% and the Post-nova Supersoft Phase

%\bibitem[Yaron et al. (2005)]{yar05}
%Yaron, O., Prialnik, D., Shara, M.M., \& Kovetz, A. 2005, \apj, 623, 398
% an extended grid of nova models II. add some model to pri95

%\bibitem[Warner (1995)]{war95}
%Warner, B. 1995, Cataclysmic variable stars, Cambridge,
%Cambridge University Press
\end{thebibliography}
\end{document}